\documentclass[fleqn,usenatbib,useAMS]{mnras}
\usepackage{newtxtext,newtxmath}

\usepackage{graphicx}	% Including figure files
\usepackage{amsmath}	% Advanced maths commands
\usepackage{multicol}        % Multi-column entries in tables
\usepackage{bm}		% Bold maths symbols, including upright Greek
\usepackage{pdflscape}	% Landscape pages

\usepackage{tikz}
\usetikzlibrary{calc, positioning}
\usepackage{ulem} % to enable strikethrough

\usepackage[percent]{overpic}

\usepackage{booktabs, multirow, makecell}

\setcellgapes{5pt}
\newcommand{\Ms}{ \mathrm{M}_\odot }
\newcommand{\kms}{ \mathrm{kms}^{-1}}

\newcommand{\LCDM}{ $\Lambda$CDM }

\title[The Local Group's mass]{The Local Group's mass: probably no more than the sum of its parts}

\author[Sawala et al.]{Till Sawala$^{1,2}$\thanks{Email: till.sawala@helsinki.fi}, Meri Teeriaho$^{1}$ and Peter~H.~Johansson$^{1}$ \\
$^{1}$Department of Physics, University of Helsinki, Gustaf H\"allstr\"omin katu 2, FI-00014 Helsinki, Finland \\
$^{2}$Institute for Computational Cosmology, Durham University, South Road, Durham DH1 3LE, United Kingdom\\
}

\date{Accepted XXX. Received YYY; in original form ZZZ}
\pubyear{2022}

\begin{document}
\label{firstpage}
\pagerange{\pageref{firstpage}--\pageref{lastpage}}
\maketitle

\begin{abstract}
The total mass of the Local Group (LG) and the masses of its primary constituents, the Milky Way and M31, are important anchors for several cosmological questions. In recent years, independent measurements have consistently yielded halo masses close to  $10^{12} \Ms$ for the MW, and $1-2 \times 10^{12}\Ms$ for M31, while estimates derived from the pair's kinematics via the `timing argument' have yielded a combined mass of around $5 \times 10^{12} \Ms$. Here, we analyse the extremely large {\sc Uchuu} simulation to constrain the mass of the Local Group and its two most massive members. First, we demonstrate the importance of selecting LG analogues whose kinematics are dominated by mutual interactions to a similar extent as the LG. Adopting the observed separation and radial velocity, we obtain a weighted posterior of $75_{-40}^{+65}$~kms$^{-1}$ for the uncertain transverse velocity. Via Gaussian process regression, we infer a total mass of $3.2^{+1.2}_{-0.9} \times 10^{12}\Ms$, significantly below the timing argument prediction. Importantly, we show that the remaining uncertainty is not rooted in the analysis or observational errors, but in the irreducible scatter in the kinematics-mass relation. We further find a mass for the less massive halo of $0.9_{-0.3}^{+0.6} \times 10^{12} \Ms$ and for the more massive halo of $2.3_{-0.9}^{+1.0} \times 10^{12}\Ms$, consistent with independent measurements of the masses of MW and M31, respectively. Incorporating the mass of the MW as an additional prior allows us to further constrain all measurements and determine that the MW is very likely to be the lower mass object of the two.
\end{abstract}


%%%%%%%%%%%%%%%%%%%%%%%%%%%%%%%%%%%%%%%%%%%%%%%%%%

%%%%%%%%%%%%%%%%% BODY OF PAPER %%%%%%%%%%%%%%%%%%

\section{Introduction}
The total mass of our own Milky Way (MW), the neighbouring Andromeda galaxy (M31), and the Local Group (LG), are key to the interpretations of many observations. For example, the abundance of massive MW satellites is consistent with \LCDM predictions only if its virial mass is not significantly higher than $\sim 10^{12}\Ms$ \citep{Boylan-Kolchin-2012, Sawala-2015}. More generally, through the predicted amount of substructure, the mass of the Milky Way also places constraints on the dark matter particle mass \citep{Lovell-2014, Kennedy-2014}. If the mass of the MW or M31 are much lower than $\sim 10^{12}\Ms$, abundance matching would imply that they are clear outliers in the galaxy-halo relation \citep{Guo-2011}. The MW also contains the best-studied `plane of satellite' \cite{Pawlowski-review}, whose significance as a challenge to the dark matter paradigm and its nature as a possibly long-lived disk, may depend both on the LG's mass assembly history \citep{Libeskind-2010}, and the depth of the potential in which the satellites evolve \citep{Sawala-2022}. Beyond satellite galaxies, the dwarf galaxies in the Local Group and their comparison to galaxies in other environments have also received attention, and the total mass of the LG is important for understanding the environment in which they form \citep[e.g.][]{Font-2022}.

Much attention has therefore been devoted in recent years to measuring the mass of the MW, M31, and the Local Group using different techniques. However, while measurements of each individual galaxy, and of the Milky Way in particular, have found masses on the order of $\sim 10^{12}\Ms$, dynamical measurements of the total mass typically yield masses of $\sim 5 \times 10^{12}\Ms$, significantly higher than the sum of both individual masses.

In this work, we use kinematics and masses of Local Group analogues from the {\sc Uchuu} simulation \citep{Ishiyama-2021}, the largest N-Body simulation of sufficient resolution, to determine the relationship between the observable kinematics and the total mass. We show how the estimates can be further constrained by assuming independent measurements for either the MW or M31, and how individual constraints, combined with the kinematics and the mass ratio distribution, can be used to infer that the mass of the MW is below that of M31. We also revisit the `Timing Argument', and explore whether there is, in fact, a discrepancy between the kinematics of the Local Group and the mass estimates for individual objects.

This paper is organised as follows. In Section~\ref{sec:previous}, we review previous measurements of the individual masses and of the total mass of the LG, highlighting the apparent discrepancy. In Section~\ref{sec:simulations}, we describe the simulations and our selection of LG analogues. In Section~\ref{sec:true}, we introduce `tidal dominance' and 'force ratio' as important criteria for linking the mass to the kinematics. We present our results for estimating the mass of the LG using Gaussian regression in Section~\ref{sec:results:regression}. In Section~\ref{sec:results:weighted}, we compute probability densities for the masses of the LG, MW, M31 and the mass ratios, based on a probability-weighted analysis. Finally,  we conclude with a summary in Section~\ref{sec:conclusion}.

\section{Previous Measurements} \label{sec:previous}
The flurry of measurements of the Milky Way mass over the past three years reflect the significance of the result, as well as the wealth of newly available data, particularly from Gaia DR2 and, most recently, Gaia DR3.

\begin{figure}
    \includegraphics[width=\columnwidth]{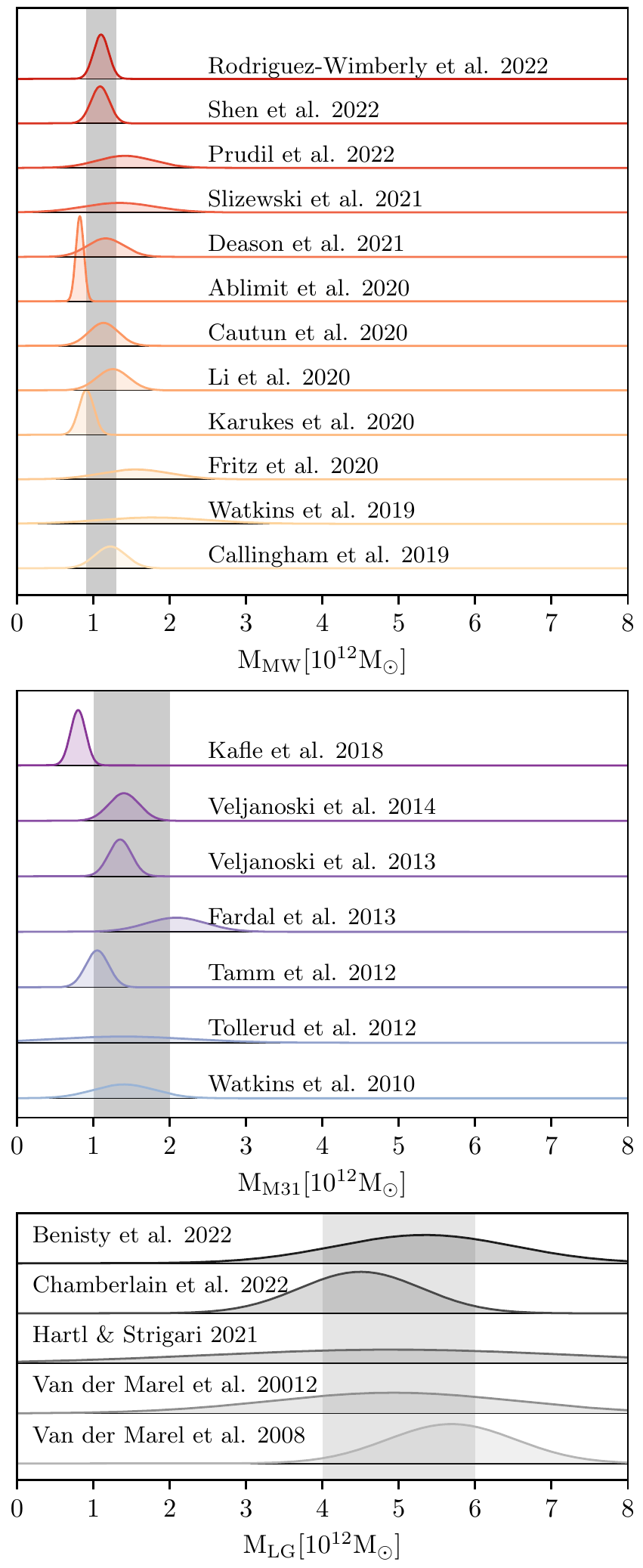}
    
    \caption{Recent direct estimates for the virial mass of the MW (top), M31 (centre), and estimates using LG kinematics for the combined mass (bottom). Each literature estimate is represented by a (skewed) normal distribution. Grey bands on the top two panels denote $\mathrm{M_{MW}} = 1.1_{-0.2}^{+0.2} \times 10^{12} \Ms$ and $\mathrm{M_{M31}} = 1.5_{-0.5}^{+0.5} \times 10^{12} \Ms$, respectively, which we use as additional constraints in some measurements. The estimates for the mass of the LG, shown in the bottom panel, are significantly higher than the sum of the masses of the two constituents.}
    \label{fig:literature}
\end{figure}

Excluding estimates that derive the mass of the MW from Local Group kinematics, recent estimates of the virial mass of the Milky Way include $1.17_{-0.15}^{+0.21} \times 10^{12} \Ms$ using Gaia DR2 satellite dynamics \citep{Callingham-2019}, $1.54_{-0.44}^{+0.75} \times 10^{12} \Ms$ using combined Gaia DR2 and HST kinematics of globular clusters \citep{Watkins-2019}, $1.51^{0.45}_{-0.40}\times 10^{12} \Ms$ using Gaia DR2 proper motions of satellite galaxies \citep{Fritz-2020} and $0.89^{+0.1}_{-0.08} \times 10^{12} \Ms$ derived from the kinematics of disk and halo stars in the galkin catalogue \citep{Karukes-2020}. 

Further mass estimates for the MW include $1.23^{+0.21}_{-0.18} \times 10^{12} \Ms$ derived from Gaia DR2 satellite kinematics and a simulation-based distribution function \citep{Li-2020}, $1.08^{+0.20}_{-0.14} \times 10^{12} \Ms$ from the Gaia DR2 rotation curve \citep{Cautun-2020}, $0.822 \pm 0.052 \times 10^{12} \Ms$ from observations of classical Cepheids \citep{Ablimit-2020}, $1.16 \pm 0.24 \times 10^{12} \Ms$ (including the mass of the LMC) from a large sample of halo stars \citep{Deason-2021}, $1.19_{-0.32}^{+0.49} \times 10^{12} \Ms$ from a Bayesian estimate using dwarf galaxy kinematics from multiple sources \citep{Slizewski-2022}, $ 1.26^{+0.40}_{-0.22} \times 10^{12} \Ms$ from high velocity RR-Lyrae stars  \citep{Prudil-2022}, $ 1.08^{+0.12}_{-0.11} \times 10^{12} \Ms$ from the H3 survey and Gaia EDR3 \citep{Shen-2022}, and finally $1.1_{-0.1}^{+0.1} \times 10^{12} \Ms$ from Gaia EDR3 proper motions of satellites when compared to \LCDM simulations \citep{Rodriguez-Wimberly-2022}. We refer the reader to \cite{Wang-2020} for a comprehensive review of earlier results. Analysis from several different tracers, and particularly from the latest studies using the most precise Gaia EDR3 observations consistently point towards a Milky Way virial mass that is very close to $10^{12} \Ms$.

\begin{figure*}
\begin{center}
    \hspace*{-4mm} \includegraphics[width=7.3in]{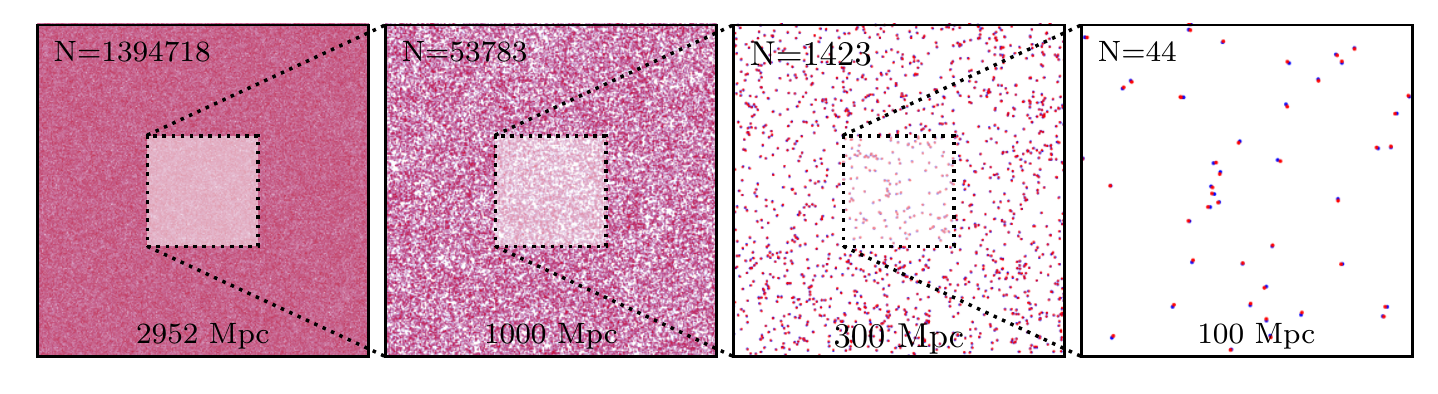}\\
\end{center}
    
    \caption{Position of Local Group analogues in the {\sc Uchuu} simulation. Blue and red dots denote the positions of the MW and M31 analogues, respectively. Each panel shows a cubic volume, the scale bar corresponds to one third of the side length. The total volume of {\sc Uchuu}, shown in the far left panel, is 25.7~Gpc$^3$ and contains $\sim 1.4$ million LG analogues. For comparison, the remaining panels 
    show zoom-ins of the full volume, with the second panel from the left comprising a 1 Gpc$^3$ volume, equal in volume to the P-Millennium simulation, the third panel from the left depicts a volume of 0.027 Gpc$^3$ equal in volume to Illustris-TNG-300, and the final panel shows a volume of 0.001~Gpc$^3$, equal in volume to DOVE / Eagle-100.
    }
    \label{fig:overview}
\end{figure*}

While M31 does not enjoy the benefit of Gaia proper motions, there exist nevertheless a number of studies of its mass. Again, excluding measurements that use LG kinematics, previous results include $1.4 \pm 0.4 \times 10^{12} \Ms$ derived from satellite kinematics \citep{Watkins-2010}, $1.2_{-0.7}^{+0.9} \times 10^{12} \Ms$ derived from kinematics of M31 dwarf spheroidals \citep{Tollerud-2012}, $\sim 1.05^{+0.15}_{-0.15} \times 10^{12} \Ms$ (combining the DM and stellar mass) from SED fitting together with the rotation curve and the kinematics of outer globular clusters and satellite galaxies \citep{Tamm-2012}, $2.0^{+0.4}_{-0.3} \times 10^{12} \Ms$ from kinematics of the Giant southern stream \citep{Fardal-2013}, $1.35_{-0.15}^{+0.15} \times 10^{12} \Ms$ \citep{Veljanoski-2013} and $1.4_{-0.2}^{+0.2} \times 10^{12} \Ms$ \citep{Veljanoski-2014}, both derived from outer halo globular clusters, and $0.8 \pm 0.1 \times 10^{12} \Ms$ estimated from high-velocity planetary nebulae \cite{Kafle-2018}. Compared to the Milky Way, the scatter in mass estimates is larger, but most estimates are in the range of $1-2 \times 10^{12} \Ms$.

Combining the best individual estimates for the masses of the MW and M31 thus leads to a total combined mass in the range of $2-3 \times 10^{12} \Ms$. These estimates also suggest a mass ratio which is close to unity, and a scenario where the mass of M31 is lower than that of the MW is certainly not ruled out. Given the difference in stellar mass, it is still commonly assumed that the mass of M31 is larger than that of the MW. However, it is worth noting that abundance matching implies higher masses for both haloes \citep{Guo-2010}, which also implies considerable scatter in the stellar-halo mass relation in order to be consistent with these mass estimates.

Meanwhile, estimates for the combined mass using the timing argument in\LCDM yield significantly higher results. Examples of direct timing argument measurements using only the radial velocity are $5.27 \times 10^{12} \Ms$  \citep{Li-2008}, $5.58_{-0.72}^{+0.85} \times 10^{12} \Ms$ \citep{vanderMarel-2008} and $4.27 _{-0.45}^{+0.45} \times 10^{12} \Ms$ \citep{vanderMarel-2012}. Accounting also for cosmic bias,  \cite{vanderMarel-2012} measure $4.93 \pm 1.6 \times 10^{12} \Ms$. Meanwhile \cite{Partridge-2013} found $4.73 \pm 1.03 \times 10^{12} \Ms$ incorporating Dark Energy into the timing argument. \cite{Benisty-2022} measure $5.6_{-1.2}^{+1.6} \times 10^{12} \Ms$ accounting for the presence of the LMC, and \cite{Hartl-2022} measure $4.75_{-2.41}^{+2.22} \times 10^{12} \Ms$. Accounting for the velocity of the MW within the LG system \citep{Petersen-2020}, \cite{Chamberlain-2022} obtain $4.5_{-0.8}^{+0.8} \times 10^{12}\Ms$ using Cepheid and Gaia data.

\begin{figure*}
    \includegraphics[width=7.1in]{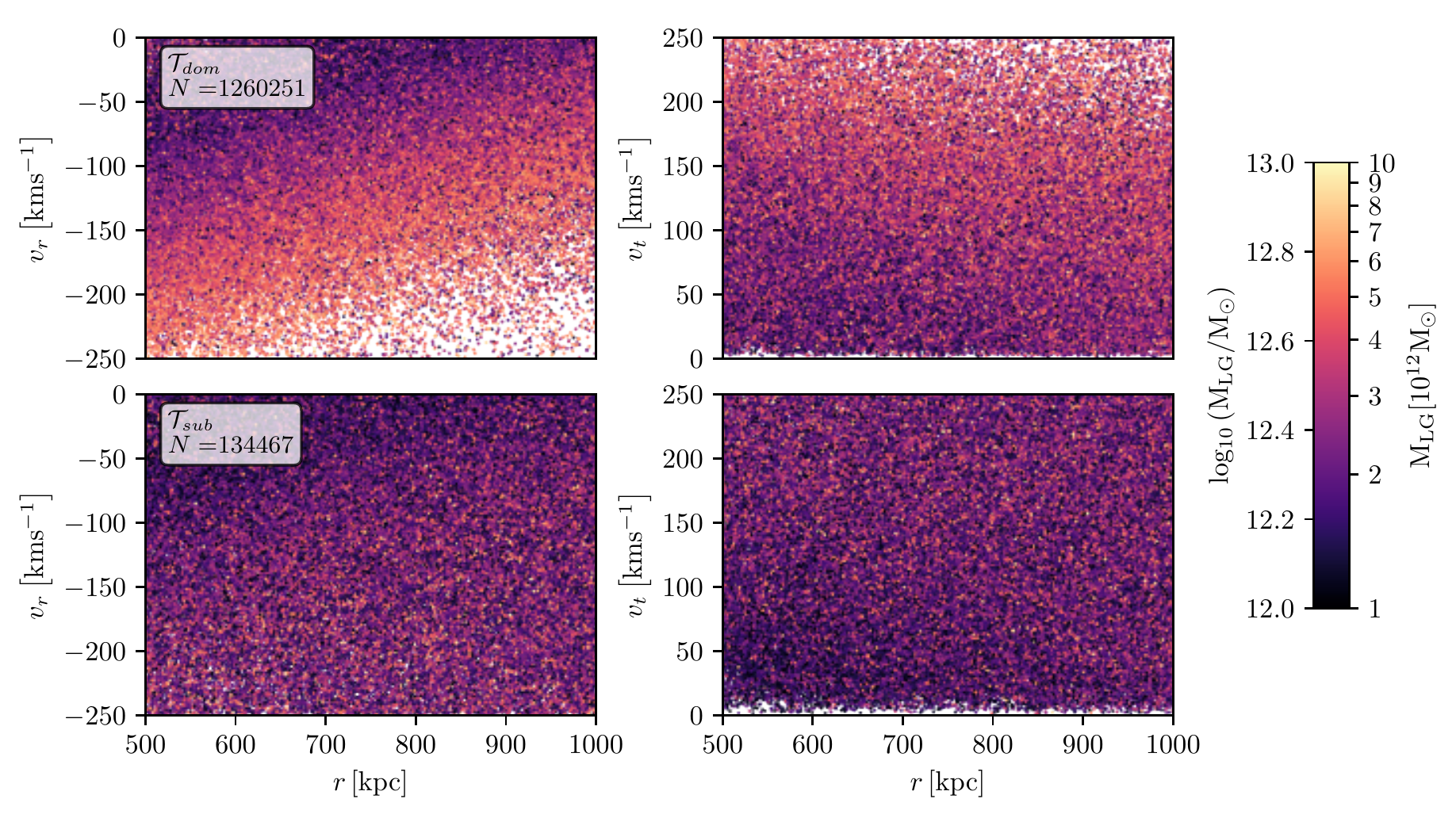}
    \vspace{-.2cm}
    \caption{The LG mass as a function of $r$ and $v_r$ (left column), and as a function of $r$ and $v_t$ (right column). The top row shows results for {\it tidally dominant} LG analogues. There is a clear correlation between mass and the kinematic parameters: at fixed separation, a greater magnitude of $v_r$ or $v_t$ both correlate with greater mass. At fixed $v_r$, a greater separation, $r$, correlates with greater mass, but at fixed $v_t$, the mass is not strongly dependent on separation. The bottom row shows results for {\it tidally subdominant} LG analogues, that is, where a third object either inside or outside the LG exerts the greatest tidal force on the lower mass halo. The number of these systems increases with separation, the masses are overall lower, and largely uncorrelated with the kinematics of the pair.
    \label{fig:correlations-tidal}
    }
\end{figure*}

\cite{McLeod-2017} used artificial Neural Networks to infer the mass of the Local Group based on LG analogues from a cosmological simulation with a volume of $\sim 0.2$ Gpc$^3$ and found a total mass of $4.9_{-1.4}^{1.3} \times 10^{12} \Ms$ when assuming a low transverse velocity of $17 \pm 17$ kms$^{-1}$ \citep{vanderMarel-2012}.

On the whole, these results are significantly higher, and also appear inconsistent with the masses of the MW and M31 measured from internal kinematic tracers, including neutral HI gas, the stellar disk, or the motions of halo stars, globular clusters, or satellite galaxies.

It is worth noting that some previous results that are not limited to pair kinematics appear to bridge the gap. \cite{Benisty-2022} report that the Timing Argument mass reduces to $3.4 \times 10^{12} \Ms$ when accounting for cosmic bias, i.e. taking into account the halo mass function. \cite{Penarrubia-2014} obtain an even lower value of $2.3 \pm 0.7 \times 10^{12} \Ms$ from a Bayesian model of galaxies extending out to a distance of 3 Mpc.

\section{Simulations} \label{sec:simulations}
This work is based entirely on publicly available data from the {\sc Uchuu} collisionless cosmological simulation \citep{Ishiyama-2021}, which samples a volume of nearly 26 Gpc$^3$ with $12800^3$ particles. In total, at $z=0$, the {\sc Uchuu} simulation contains $\sim 250$ million haloes in the mass range of $0.5 - 5 \times 10^{12} \Ms$, and a particle mass of $4.83 \times 10^8 \Ms$ implies that the lowest mass haloes we consider are resolved with more than 1000 particles. {\sc Uchuu} uses cosmological parameters corresponding to the Planck-2018 results, namely, $\Omega_0=0.3089$, $\Omega_{\rm b}=0.0486$, $\Omega_{\Lambda} =0.6911$, $h=0.6774$, $n_s=0.9667$, and $\sigma_8=0.8159$. The simulation outputs have been processed using the {\sc Rockstar} phase-space structure finder. All results reported in this paper are expressed in terms of physical coordinates, distances in kpc, masses in solar masses ($\Ms$), and velocities in $\kms$. For the halo mass, we use the most common definition $M_{200,c}$, i.e. the mass contained within a sphere enclosing a mean density of 200 times the critical density.

\subsection{Definition of Local Group analogues} \label{sec:selection}
Observations accurately measure the distance between the centres of the MW and Andromeda as $r=765 \pm 11$~kpc \citep{Chamberlain-2022}, and the radial velocity as $v_r=109.3 \pm 4.4$~kms$^{-1}$ \citep{vanderMarel-2012}. The transverse velocity is less certainly known, and some measurements strongly differ. Using only HST proper motions, \citep{vanderMarel-2012} found $v_t = 17 \pm 17$kms$^{-1}$, which would imply an extremely radial orbit. Later combining HST and Gaia DR2 proper motions, \cite{vanderMarel-2019} find a somewhat higher value and measure $v_t = 57^{+35}_{-31}$kms$^{-1}$. This measurement constitutes a weighted average of an HST-only value of $36^{+39}_{-26}$kms$^{-1}$, and a Gaia-only value of $133^{+70}_{-68}$kms$^{-1}$. \cite{Salomon-2016} derived an even greater transverse velocity of $v_t = 164.4 \pm 61.8$kms$^{-1}$ by studying the projected line-of-sight velocities of M31 satellites, and comparing the results to those found in simulations. \cite{Salomon-2020} measure $v_t = 82.4 \pm 31.2$kms$^{-1}$ using Gaia DR3 proper motions of blue main sequence stars in M31. We adopt here $v_t = 57 \pm 35 $kms$^{-1}$, very similar to the combined result of \cite{vanderMarel-2019}. Whenever we Monte-Carlo sample the uncertainties, or construct probability-weighted estimates, we assume that the probability distributions for the true values of $r$, $v_r$ and $v_t$ are Gaussian, subject to the additional constraints $r >0 $ and $v_t > 0$.

Our broadest category of Local Group analogues is defined as a pair of haloes that satisfy all of the following criteria:

\begin{itemize}
    \item each M$_{200}$ mass within $[0.5 \dots 5] \times 10^{12} \Ms$
    \item separation, $r$, within $[500 \dots 1000]$ kpc
    \item radial velocity, $v_r$, within $[-250 \dots 0] \kms$
    \item transverse velocity, $v_t$, within $[0 \dots 250] \kms$
    \item no further halo more massive than $0.5\times 10^{12} \Ms$ within 2 Mpc
\end{itemize}

It is worth noting that the kinematics constrain primarily the total mass of the LG. The kinematics also allow, for example, a high mass halo paired with a halo below the defined mass range, so we must consider the lower limit of $0.5 \times 10^{12} \Ms$ for each halo as an additional, fixed prior. As we will discuss, this has implications for the inferred mass ratio, but is, however, well-motivated from other observations.

In total, the {\sc Uchuu} simulation contains $\sim 1.4$ million systems defined using the criteria above. In Figure~\ref{fig:overview}, we show the distribution of these LG analogues within the volume. The panel on the left shows the entire volume of nearly 26 Gpc$^3$. The uniquely large volume of the {\sc Uchuu} simulation at this resolution allows us to retain sufficiently large sample sizes, even when additional constraints are imposed. 

It should also be noted that the Local Group, whether in the real universe or in cosmological simulations, is not a true two-body system. It includes mass beyond the respective $r_{200}$ of its two main haloes, to which different tracers may be sensitive to different degrees \citep{Penarrubia-2017}. Throughout this paper, we adopt the convention of defining the mass of the Local group as the sum of the two halo masses, or $\mathrm{M_{LG}} = \mathrm{M_1} + \mathrm{M_2}$. We distinguish between the more massive halo whose mass we label $\mathrm{M_1}$ and the less massive halo of mass $M_2$. We do not generally associate them with the MW or M31 specifically, except where we additionally constrain the MW halo mass in Section~\ref{sec:results:MW}, where we label the mass of the MW as $\mathrm{M_{MW}}$ and that of M31 as $\mathrm{M_{31}}$.

\begin{figure*}
    \includegraphics[width=7.1in]{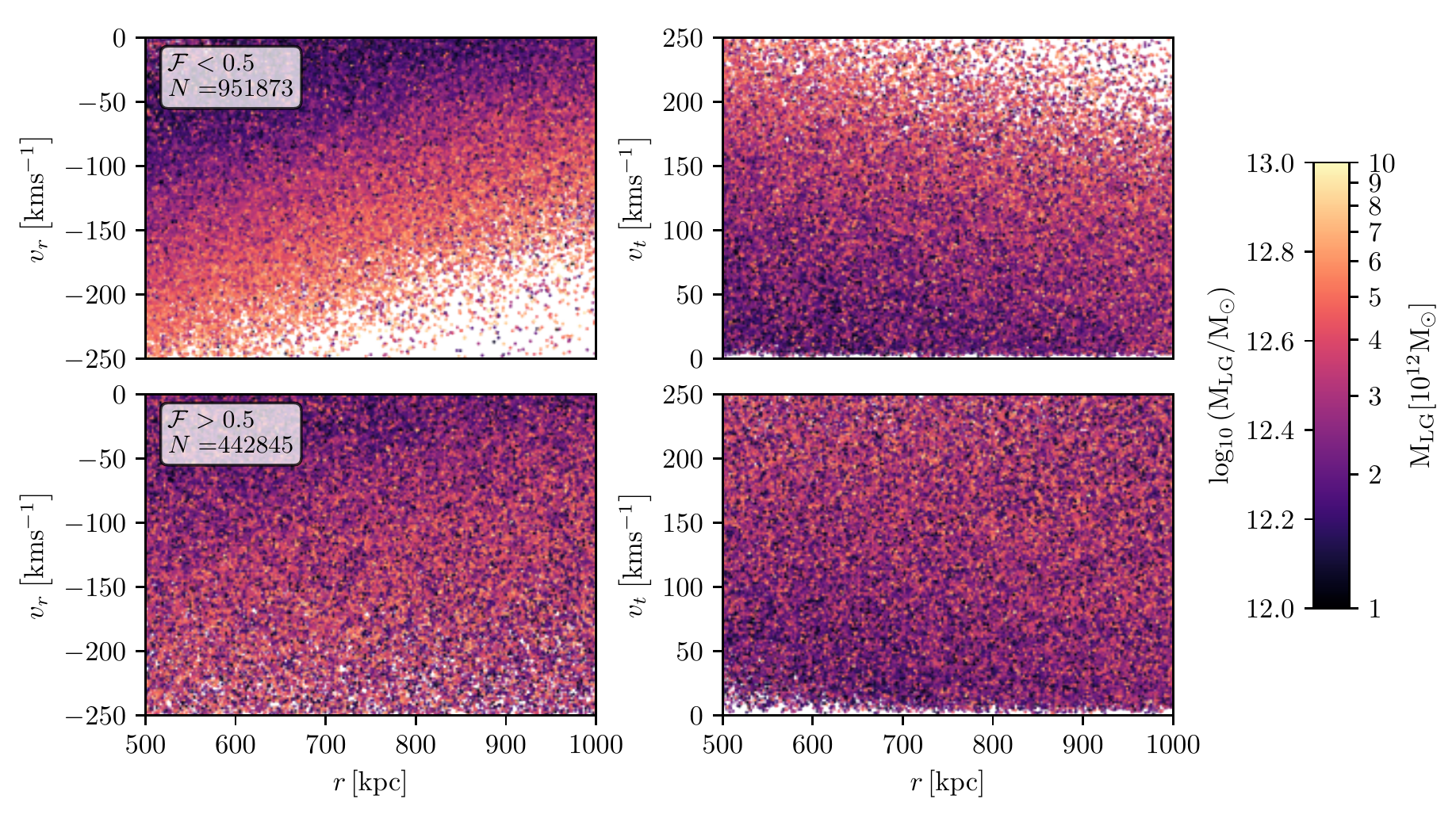}
    \vspace{-.2cm}
    \caption{The LG mass as a function of $r$ and $v_r$ (left column), and as a function of $r$ and $v_t$ (right column), similar to Figure~\ref{fig:correlations-tidal}. The top row shows results for low force ratio systems ($\mathcal{F} < 0.5$), while the bottom row shows high force ratio systems ($\mathcal{F} > 0.5$). In low force ratio systems, the gravitational fields of the two haloes dominate their overall gravitational interactions, resulting in a clear correlation between kinematics and mass. In high force ratio systems, other haloes cause significant perturbations to the gravitational field, blurring this correlation. The increased scatter with increased $\mathcal{F}$ also means that regions only sparsely populated in the top panels, such as very low $v_r$ or very high $v_t$, contain significantly more systems in the bottom panels.
    }
    \label{fig:correlations-force}
\end{figure*}

\section{True Local Group analogues} \label{sec:true}
The degree to which the mass of the LG is related to the kinematics of the MW-M31 orbit depends on the degree to which the motion of each object is affected by their common gravitational field, or by other haloes. For example, the `Timing argument' assumes that the system evolves in complete isolation. Previous works \citep[e.g.][]{Li-2008, Fattahi-2015, Carlesi-2016} that selected LG analogues from cosmological simulations have typically excluded systems that contain a third object within some radius, either above the lower limit of the range of halo masses considered or, less restrictively, more massive than the less massive of the pair.

However, we find that this is not sufficient to ensure that the dynamics of the simulated LG analogues are free of undue outside influences, i.e. greater than assumed to be the case for the real LG. In this section, we introduce two new criteria, {\it tidal dominance}, and the external-to-internal {\it force ratio}. We use them together to identify {\it true} LG pairs, i.e. those pairs governed by internal kinematics to a similar extent as the real Local Group.

\subsection{Tidal dominance}
The {\sc Uchuu} halo catalogues contain information about the hierarchy of tidal forces. For each halo, we can identify a more massive halo that exerts the greatest tidal force. This allows us to identify LG analogues in which the tidal field of one LG member is dominated by the other, as opposed to those for which the tidal fields of both haloes are dominated by other objects. We refer to LG analogues in the first category as `tidally dominant' and the remainder as `tidally subdominant'.

It is expected that the relations between kinematics and mass only hold for tidally dominant systems. Indeed, this is the case. In Figure~\ref{fig:correlations-tidal}, we plot distributions of {\sc Uchuu} LG analogue masses as functions of the three kinematic parameters. The left panel shows total mass, $\mathrm{M_{LG}}$, as a function of $r$ and $v_r$, while the right panel shows $\mathrm{M_{LG}}$ as a function of $r$ and $v_t$. For {\it tidally dominant} systems, shown on top, there are clear correlations, as also shown in previous studies \citep[e.g.][]{Fattahi-2015}: at fixed separation, LG analogues with more negative radial velocity or higher transverse velocity tend to have higher masses, although some scatter is also apparent.

For {\it tidally subdominant} systems, shown on the bottom and comprising $\sim 8\%$ of the total sample, the kinematics are primarily governed by forces outside the system itself. Consequently, the mass of the system is largely uncorrelated with the kinematics. In particular, we find a large number of low-mass tidally subdominant systems with kinematics similar to the LG not due to internal processes, but due to random, external perturbations.

We refer to the subsets of tidally dominant and subdominant LG analogues  as $\mathcal{T}_{dom}$ and $\mathcal{T}_{sub}$, respectively, and remove the subdominant systems from most of our subsequent analysis.

\subsection{Force Ratio}
We also consider the gravitational forces exerted by the two LG members on each other, and compare them to the forces exerted on each member by every other halo above $2\times 10^{11} \Ms$ within 5 Mpc of the LG centre. We parameterise the ratio between the maximum force exerted onto either LG member by any of these haloes, and the force exerted by the other member, as the force ratio:

\begin{equation}
    \mathcal{F} = \mathrm{max} \left( \frac{\mathrm{max}(M_i / r_{1i}^2)}{M_2 / r^2}, \frac{\mathrm{max}(M_k / r_{2j}^2)}{M_1 / r^2} \right),
\end{equation}

where $\mathrm{M}_i$ and $r_{1i}$ are the mass and distance of a halo from one member of the pair, and $\mathrm{M}_j$ and $r_{2j}$ are the mass and distance of a halo from the other member. A low value of $\mathcal{F}$ means that the forces which the two LG members exert on each other dominate the gravitational interactions of both haloes, while a high value of $\mathcal{F}$ means that, for at least one of the two  LG members, at least one other halo exerts a significant force.

\begin{figure*}
    \includegraphics[width=7.1in]{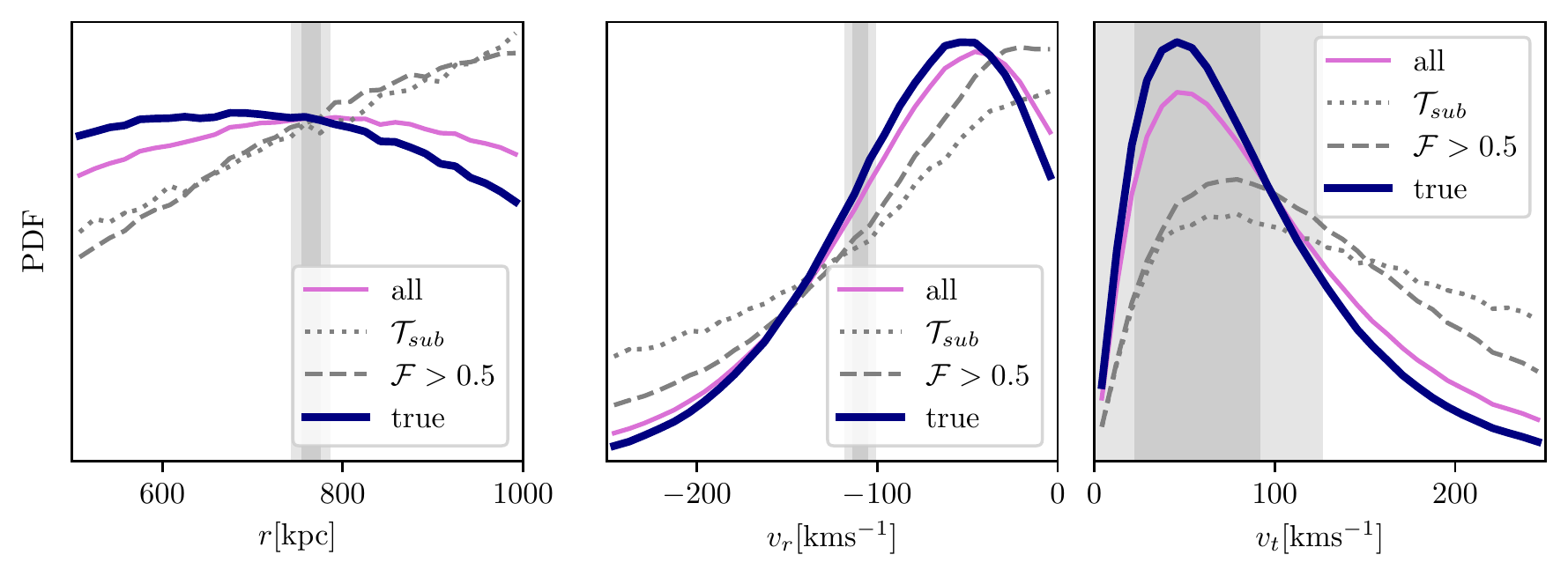}
    \vspace{-.2cm}
    \caption{Distributions of observables, $r$ (left), $v_r$ (centre) and $v_t$ (right) in our LG analogues. On all panels, shaded vertical bands indicate $\pm 1 \sigma$ and $\pm 2 \sigma$ of the observations. Pink lines show the full sample of LG analogues that match the kinematic criteria. Grey lines show LG analogues we exclude from our analysis: for $\mathcal{T}_{sub}$ (dotted lines), none of the LG members dominate the other's tidal force, while for $\mathcal{F} > 0.5$ (dashed lines), a third halo exerts a gravitational force on one of the LG members which is larger than 0.5 of the force exerted by the other member. Both of these subsets contain haloes with very high or low radial, and high transverse velocities, which are not related to the internal kinematics. When haloes in either of these two subsets are removed, the remainder ('true pairs', thick blue lines) is more likely to match each of the observed values, and shows a stronger correlation between mass and kinematics.
    }
    \label{fig:distribution-observables}
\end{figure*}

The kinematics of systems with low force ratios are determined by the LG itself, while the kinematics of systems with high force ratios are strongly affected by haloes outside the pair. We classify our LG analogues according to  $\mathcal{F}$ as either ``internally dominated´´, with $\mathcal{F} < 0.5$ comprising $\sim 2/3$ of all systems, and ``significantly externally affected´´ with $\mathcal{F} \geq 0.5$, comprising the remaining $\sim 1/3$ of the sample. 

In Figure~\ref{fig:correlations-force}, we show the correlations of mass and kinematics for LG analogues in the three categories. We find that there is a clear separation: halo pairs with low force ratios have the strongest correlation with the least amount of scatter, while halo pairs with high force ratios show a weaker correlation between mass and kinematics, and significantly more scatter.
The real LG most likely contains no halo with a mass above $2\times 10^{11}\Ms$. With a distance to Centaurus A of $3.4$~Mpc, approximately $5\times$ the distance between the MW and M31, and its mass approximately $10\times$ greater than that of the MW or M31, we estimate $\mathcal{F}_{LG} \sim 0.4$.

Throughout the rest of this paper, we will refer to the set of LG analogues that are both `tidally dominant' and have force ratios $\mathcal{F} < 0.5$, as {\it true} pairs.

\subsection{Distributions of kinematic variables} \label{sec:results:kinematics}
In Figure~\ref{fig:distribution-observables}, we show the distributions of the three kinematic variables, $r$, $v_r$ and $v_t$. On all panels, grey bands indicate the assumed $1 \sigma$ and $2 \sigma$ observational errors, as discussed in Section~\ref{sec:selection}. Pink solid lines show the distributions for the full set of LG analogues, thick blue solid lines show the distributions for the `true' LG analogues. Grey lines show the distributions for the LG analogues not included in the true sample: dotted for tidally subdominant systems and dashed for those with $\mathcal{F} \geq 0.5$.

It can be seen that all three variables are distributed differently for the different subsets. True LG analogues have, on average, smaller separation than those in the $\mathcal{F} \geq 0.5$ or $\mathcal{T}_{sub}$ subsets, and smaller than the full sample, where these two subsets are included. This is expected because, all else being equal, the internal forces are relatively stronger for lower separation systems. The fraction of systems that match the observations for $r$ is nearly identical in all subsets.

The differences in radial and transverse velocity are more pronounced: the set of true LG analogues contain significantly more objects with either very negative $\bigl(v_r < - 150 \mathrm{kms}^{-1}\bigr)$ or very small $\bigl(v_r > -50 \mathrm{kms}^{-1}\bigr)$ radial velocities, or with high $\bigl(v_t > 100\mathrm{kms}^{-1}\bigr)$ transverse velocities, compared to the $\mathcal{F} \geq 0.5$ and $\mathcal{T}_{sub}$ subsets. In fact, the majority of the full sample of LG analogues that meet our selection criteria, and that have either very negative radial velocities or very high transverse velocities, are not `true' LG analogues governed by mutual interactions, but are systems whose kinematics are caused by other objects. It is worth noting that removing these objects causes the `true' LG analogues to have a significantly higher fraction of objects within the observational uncertainties.

\subsection{The transverse velocity}
Interestingly, \cite{Forero-Romero-2022} recently reported that the average transverse velocity of LG analogues is higher in simulations with large box sizes, pointing to a limitation of studying LG analogues in relatively small volume simulations. The largest simulation they consider, {\sc Abacus Summit}, has a box size almost identical to {\sc Uchuu}, albeit at 8 times lower mass resolution. Interestingly, they find a transverse velocity with a median and $1\sigma$-equivalent uncertainty of $105^{+94}_{-59}$~kms$^{-1}$, significantly higher than our results: $v_t = 81_{-46}^{+74}$~kms$^{-1}$ for all LG analogues, and $v_t = 72_{-40}^{+65}$~kms$^{-1}$ for true pairs. If we further incorporate the observed values and uncertainties on $r$ and $v_r$ as priors, we obtain a nearly identical weighted posterior of $v_t = 75_{-40}^{+65}$~kms$^{-1}$ for the LG transverse velocity, consistent with the value of $v_t = 57^{+35}_{-31}$kms$^{-1}$ of \cite{vanderMarel-2019}.

The difference between our result and those of \cite{Forero-Romero-2022} could be partly due to the fact that their selection of haloes with $v_\mathrm{max} = 200 \dots 260$~kms$^{-1}$ led them to consider more massive LG analogues which, as we will discuss in Section~\ref{sec:results:inverse}, tend to have higher transverse velocities. Perhaps more significantly, due to the fact that they include pairs with separations up to 1.5 Mpc, their sample almost certainly contains a large fraction of unbound pairs, or pairs whose kinematics are not determined internally, but through interactions with other haloes. For example, we find $v_t = 118_{-68}^{+82}$~kms$^{-1}$ for tidally subdominant pairs, and  $v_t = 107_{-59}^{+78}$~kms$^{-1}$ for pairs with high force ratios. Both of these are close to the value obtained by \cite{Forero-Romero-2022}. However, they may not be a good prior for the transverse velocity of the Local Group.

\begin{figure*}
    \includegraphics[width=7.1in]
    {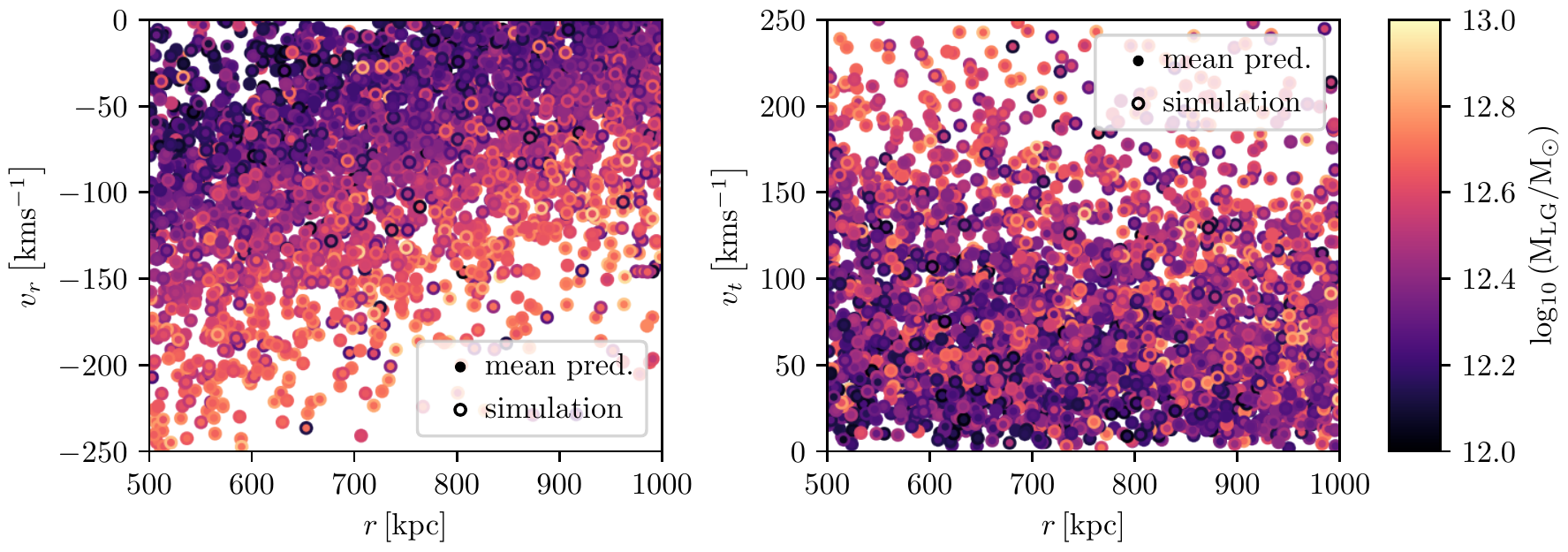}
    \vspace{-.3cm}
    \caption{Local Group mass measured in the simulation (symbol edges) and mean predicted mass using Gaussian regression (symbol centres) as a function of the observed kinematics, for a random subset of $\mathcal{T}_{dom}, \mathcal{F} < 0.5$ LG analogues. Similar colours for the centre and edge mean that the regression predicted a mass similar to that in the simulation. The regression correctly captures both the overall mass gradients detectable by eye, as well as, in many cases, the masses of apparent outliers.}
    \label{fig:regression-kinematics}
\end{figure*}

\begin{figure}
   \includegraphics[width=\columnwidth]{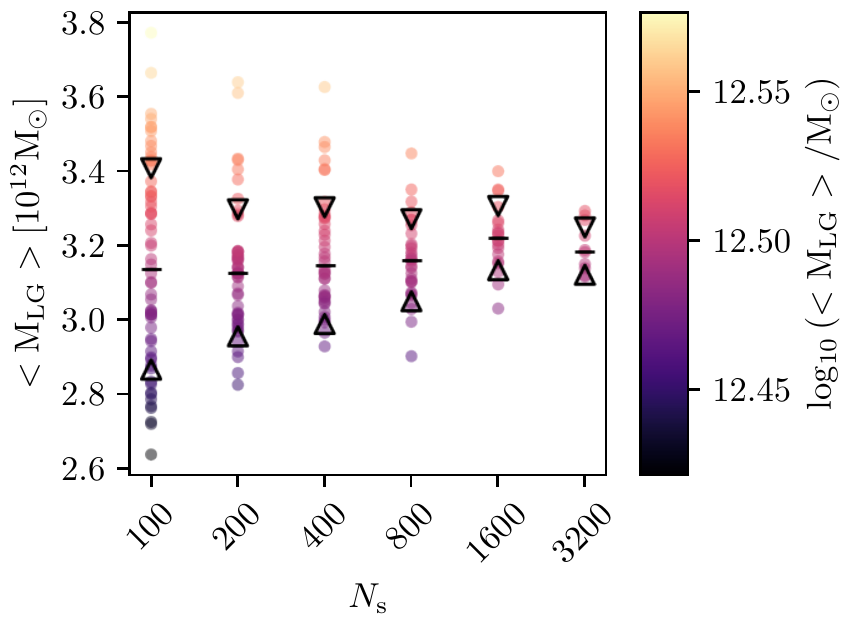}
   \vspace{-.4cm} 
    \caption{Individual predicted median values of $\mathrm{M_{LG}}$ from independent Gaussian process regressors trained on $N_\mathrm{s}$ data points, after Monte-Carlo sampling the observational uncertainty. Black horizontal lines indicate the median of the values plotted for each sample size, upwards and downwards triangles indicate $\pm 1 \sigma$ scatter. As the number of points in the sample increases, the scatter in the predictions is reduced. At $N_\mathrm{s}=3200$, we obtain an expectation value for the median of $\mathrm{M_{LG}} = 3.18 \pm 0.06 \times 10^{12} \Ms$.}
    \label{fig:regression-convergence}
\end{figure}

\begin{figure}
    \includegraphics[width=\columnwidth]{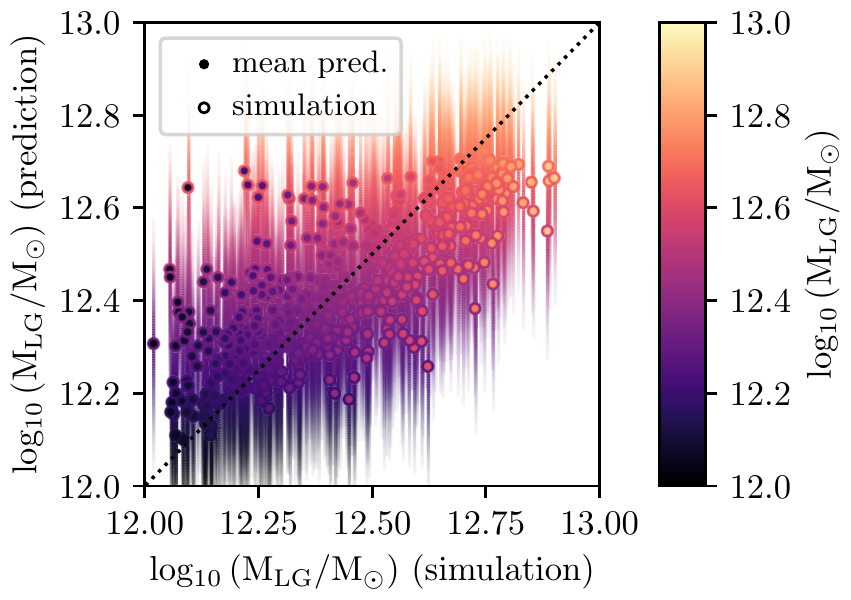}
    \vspace{-.3cm}
    \caption{The LG mass predicted using Gaussian regression ($y$-axis) compared to the mass measured in the simulation ($x$-axis). As in Figure~\ref{fig:regression-kinematics}, symbol centres are coloured according to the mean prediction, symbol edges are coloured according to the mass in the simulation. Error bars extend to $\pm 1 \sigma$ of the uncertainty of each prediction. There is a clear correlation between predicted and measured mass, with no apparent bias. Some amount of regression-to-the-mean is also apparent: for low masses, the Gaussian regression tends to overpredict the mass, while for high masses, it underpredicts it. Consequently, the range of mean predictions is narrower than the full range of masses in the simulations. However, this apparent shortfall can be rectified when the uncertainty of the prediction is taken into account.}
    \label{fig:regression-error}
\end{figure}

\section{The LG mass via Gaussian Process Regression} \label{sec:results:regression}
We estimate the mass of the LG, $\mathrm{M_{LG} = M_{MW}+M_{M31}}$, using Gaussian process regression (GPR), a machine learning algorithm that provides an estimate of both the most likely value, and the uncertainty (standard deviation) of the estimate. Specifically, we infer the relationship between the inputs (kinematics) and the target (Local Group mass), i.e. the conditional mass distribution given the observed kinematics.

% This section still needs a bit of work...
GPR is a type of supervised machine learning, meaning that it first ‘learns’ the connection between the input and the target variables from a given dataset where both are known. In our case, the regressor learns the correlation between the kinematics (input) and the LG mass (target) from sets of `true' LG analogues from the {\sc Uchuu} simulation. Afterwards, the trained regressor can predict the LG mass from a given new set of kinematics. During the training, functions are fitted to the training data, whose coefficients are optimised iteratively. Without further constraints, there is an infinite number of possible functions that can fit a given set of data. This sets GPR, as a form of non-parametric regression, apart from parametric regression models, where the functional form is fixed, and only function parameters are optimised.

\begin{figure}
    \includegraphics[width=\columnwidth]{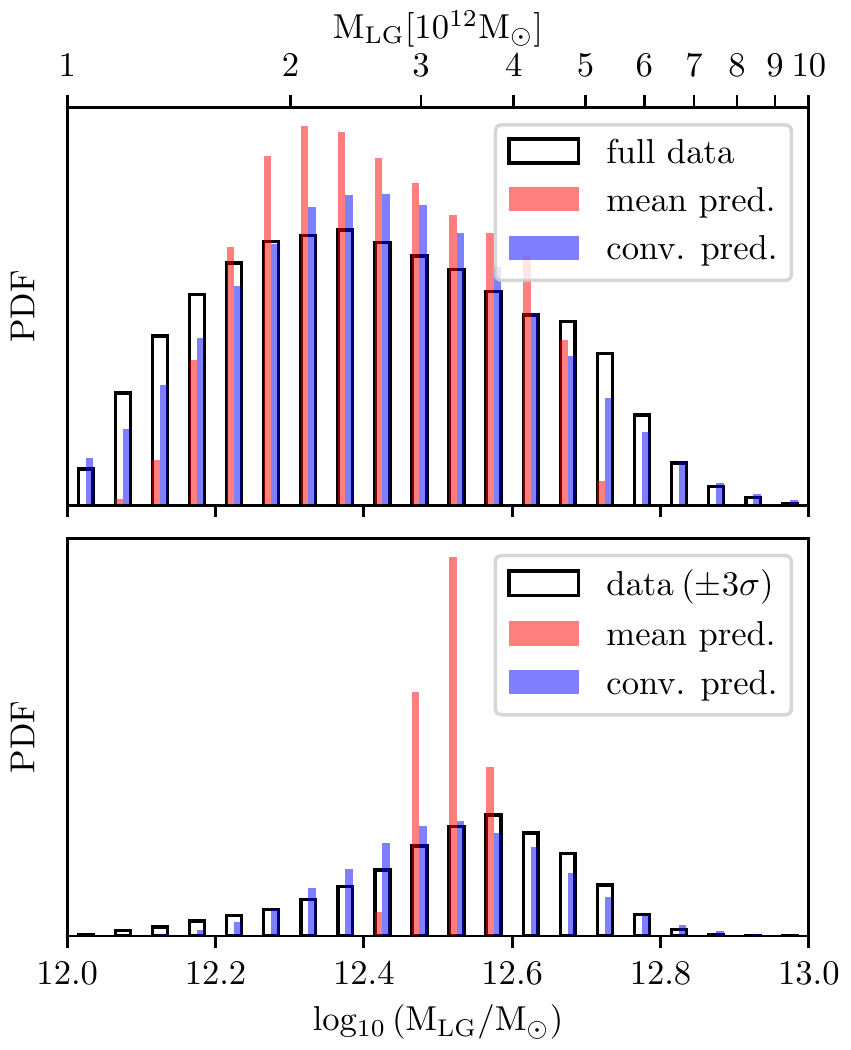}
    \vspace{-.3cm}
    \caption{Distribution of LG masses in the simulation (black open histograms), predicted means of the Gaussian Process regression (red histograms), and after the predicted means are convolved with the respective predicted uncertainties (blue histograms). The top panel shows the results for all LG analogues in the tidally dominant, $\mathcal{F} < 0.5$ sample. The bottom panel shows the results for only those LG analogues that are within $\pm 3 \sigma$ for each of the kinematic variables on which the prediction is based, $r$, $v_r$ and $v_t$. It can be seen that the distributions of mean predictions are more concentrated than those of the simulation data, lacking both high- and low mass outliers. By contrast, after convolving the predicted means with the predicted uncertainties, the distribution of the data is reproduced well. This effect is particularly pronounced when, as shown in the bottom panel, the range of input variables is narrowed, in this case to $\pm 3 \sigma$, where $\sigma$ is the default observational uncertainty.}
    \label{fig:regression-histogram}
\end{figure}

The goal of the regression is to find a probability distribution of all functions that fit the data, and from that, predict both the mean and variance of the mass for a given set of kinematics, over the entire range of possible input values. In our case, we construct the possible functions as the dot products (inner product) of a radial basis function (RBF), a constant, and a white noise term. The regression begins with a uniform prior on the probabilities of all possible functions. Subsequently, the probability distribution is successively updated based on new information in the form of training data. The posterior distribution of the possible function is then used to predict both the mean and the variance (or standard deviation) \citep{Wang-2020-ML}. In this work, we use the {\sc scikit-learn} \citep{scikit-learn} implementation of GPR, {\sc GaussianProcessRegressor}.

Figure~\ref{fig:regression-kinematics} shows the mean prediction of the LG mass obtained via Gaussian Process regression (symbol centres) in the $r-v_r$ plane (left panel), and in the $r-v_t$ plane (right panel), compared to the values measured for the same systems in the simulation (symbol edges). It can be seen that the Gaussian Process regression predicts the gradients in mass identifiable by eye, but also, in many cases, correctly predicts the masses of points that appear as outliers in each individual 2D plane, due to the fact that it uses the full, 3D kinematic information in its estimate. However, as demonstrated by the data points with differing edge- and centre colours, the regression is not perfect.

\subsection{Predicted Uncertainty}
The algorithmic complexity of GPR of $\mathcal{O}(N^3)$ prohibits us from using the full data for training. However, GPR is quite efficient even for small sample sizes. In Figure~\ref{fig:regression-convergence}, we show the convergence behaviour of the regression. We train the regressor multiple times on $N_\mathrm{s}$ data points, and for each instance, we obtain an independent estimate of the LG mass via Monte-Carlo sampling of the observations (see Section ~\ref{sec:MC-sampling}). The precision of the regression increases with the number of data points: for a sample size of $N_\mathrm{s} = 100$, we find $\mathrm{M_{LG}} = 3.14 \pm 0.27 \times 10^{12} \Ms$, while for a sample size of $N_\mathrm{s} = 3200$, we find a median value of $\mathrm{M_{LG}} = 3.18 \pm 0.06 \times 10^{12} \Ms$. The uncertainty of $\pm 0.06 \times 10^{12} \Ms$ represents the uncertainty of our prediction of the {\it median} value of the LG mass distribution. As we explain below, however, it is quite different from the uncertainty of our prediction of the mass of the LG.

In Figure~\ref{fig:regression-error} we plot the predicted mass as a function of mass in the simulation, with symbols coloured, as in Figure~\ref{fig:regression-histogram}, according to the mass in the simulation in the symbol centres, and according to the mean prediction at the symbol edges. It can be seen that the range of mean predictions is narrower than the range in the simulations: the ensemble of mean predictions lacks both the high-and low mass outliers found in the simulation. The mean prediction underpredicts the mass of the most massive objects, and overpredicts the masses for the least massive ones. This behaviour is not unexpected: indeed, if the connection between kinematics and mass contains some amount of random scatter, the extremely low and high mass points are likely to fall, respectively, on the low and high mass side of the scatter. While the mean regression cannot, by definition, reproduce the random behaviour for individual objects, we can account for the presence of scatter by including a white noise term in the kernel function.

\begin{figure}
    \includegraphics[width=\columnwidth]{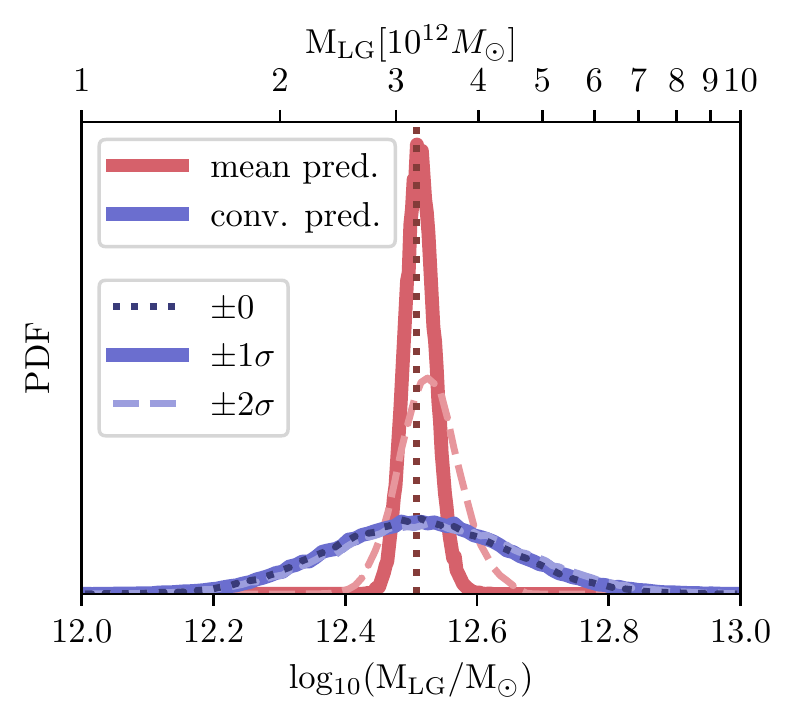}
    \vspace{-.4cm}
    \caption{Local Group mass predicted by the Gaussian regression, when using Monte Carlo samples of the observed kinematics within the stated uncertainties as inputs. The red lines show the distribution of the mean predictions, the blue lines show the distribution after convolving each mean prediction with its associated uncertainty. It can be seen that the mean prediction appears to be very precise, but this belies the true uncertainty in the LG mass-kinematics relation. Only after convolving with the uncertainties, we get a realistic estimate, and find a mass of $3.2^{+1.2}_{-0.9} \times 10^{12}\Ms$ when assuming standard $\pm 1 \sigma$ observational errors. While the uncertainty of the mean prediction depends strongly on the assumed observational errors (dotted lines: zero error, solid lines: the default errors, dashed lines: twice the default errors), the uncertainty of the convolved prediction is dominated by the uncertainty in the kinematics-mass-relation.}
    \label{fig:regression-MC}
\end{figure}

One key advantage of the Gaussian regression over some other regression models (such as decision trees or neural networks) is that it automatically provides an estimate of the uncertainty, or error, of each estimate along with the mean estimate. We show the predicted $1\sigma$-uncertainties for each data points as vertical lines in Figure~\ref{fig:regression-error}. The range of values within the uncertainties around the mean predictions shown in Figure~\ref{fig:regression-error} extends beyond that of the mean values alone. For a full prediction that encapsulates both the predicted trend and the irreducible uncertainty, we therefore convolve each predicted mean with a normal distribution of the predicted standard deviations.

In Figure~\ref{fig:regression-histogram}, we compare the distribution of LG masses measured directly in the simulation with those obtained by applying the Gaussian regression to the kinematics measured in the simulations for the same objects. On the top panel, the open histogram shows the full data in the tidally dominant, $\mathcal{F} < 0.5$ sample, i.e. the `true' LG analogues. The red histogram shows the distribution of mean predictions of the regression, while the blue histogram shows the distribution after convolution of the predicted means with the predicted uncertainties. As discussed above, the mean predictions do not fully account for the scatter, but after the uncertainty is taken into account, the resulting distribution reproduces the shape of the data. On the bottom panel, the data is reduced to the subset for which each observable, $r$, $v_r$ and $v_t$, is limited to $\pm 3 \sigma$ around the observational value, where $\sigma$ is the quoted observational error. Here, the difference between the distribution of masses in the data and in the distribution of mean predictions, is even more pronounced: while the distribution of the mean predictions has $\mathrm{M_{LG}} = 3.4^{+0.3}_{-0.2}\times 10^{12} \Ms$, the data has $\mathrm{M_{LG}} = 3.5^{+1.2}_{-1.1}\times 10^{12}$, a similar median, but much greater scatter. After convolution with the predicted uncertainty, the prediction has $\mathrm{M_{LG}}=3.4^{+1.3}_{-0.9} \times 10^{12} \Ms$, reproducing well both the median and the scatter in the data.

\subsection{A prediction of the LG mass} \label{sec:MC-sampling}
To predict the mass of the actual Local Group based on our regression, we use the observed values of $r$, $v_r$ and $v_t$ as inputs to the regressor trained on the simulation data. In Figure~\ref{fig:regression-MC}, we show the resulting probability density for the LG mass. To account for the observational uncertainty, we create Monte-Carlo samples for $10^6$ combinations of the observables, each following a normal distribution with mean and standard deviations inferred from the observations, as described in Section~\ref{sec:selection}. We truncate the Monte Carlo samples to exclude nonsensical values, i.e. $r < 0$ or $v_t < 0$, but this has only a minor effect on the results.

The thick red line in Figure~\ref{fig:regression-MC} shows the resulting mass distribution using the mean values of the regressor with the quoted observational uncertainties. The predicted mass is $3.43^{+0.1}_{-0.2}\times 10^{12} \Ms$ - an apparently impressively precise result. As indicated by the dashed and dotted red lines, the precision is limited only by the observational errors: the uncertainty increases to $\pm 0.3 \times 10^{12} \Ms$ when the observational errors are assumed to be twice as high (dashed red line) and becomes infinitesimally small when the assumed errors go to zero (dotted vertical line). Notably, as we will show in Section ~\ref{sec:results:weighted}, this behaviour is similar to an analysis based on the classical 'timing argument'. However, it is not a true reflection of the uncertainty of the estimate of the mass based on the kinematics.

The blue lines in Figure~\ref{fig:regression-MC} show, for the same Monte Carlo samples with the same assumed observational uncertainties, the LG mass distribution after the predicted means are convolved with the predicted uncertainties. These estimates account both for the observational errors and for the uncertainty in the relation between kinematics and mass. With the standard, $\pm 1 \sigma$ observational errors, we obtain a mass of $\mathrm{M_{LG}} = 3.2^{+1.2}_{-0.9} \times 10^{12} \Ms$ - a similar mass with a larger, but more realistic estimated uncertainty. Comparing this result to the one shown by the red lines, and discussed above, it is clear that the precision of estimating the LG mass is not limited by the current observational error, but by the fundamental scatter in the relation between the kinematics and mass. Indeed, when we bring the assumed observational error to zero, as shown by the dotted blue line, the mass estimate hardly changes.

\begin{figure*}
    \includegraphics[width=7.1in]{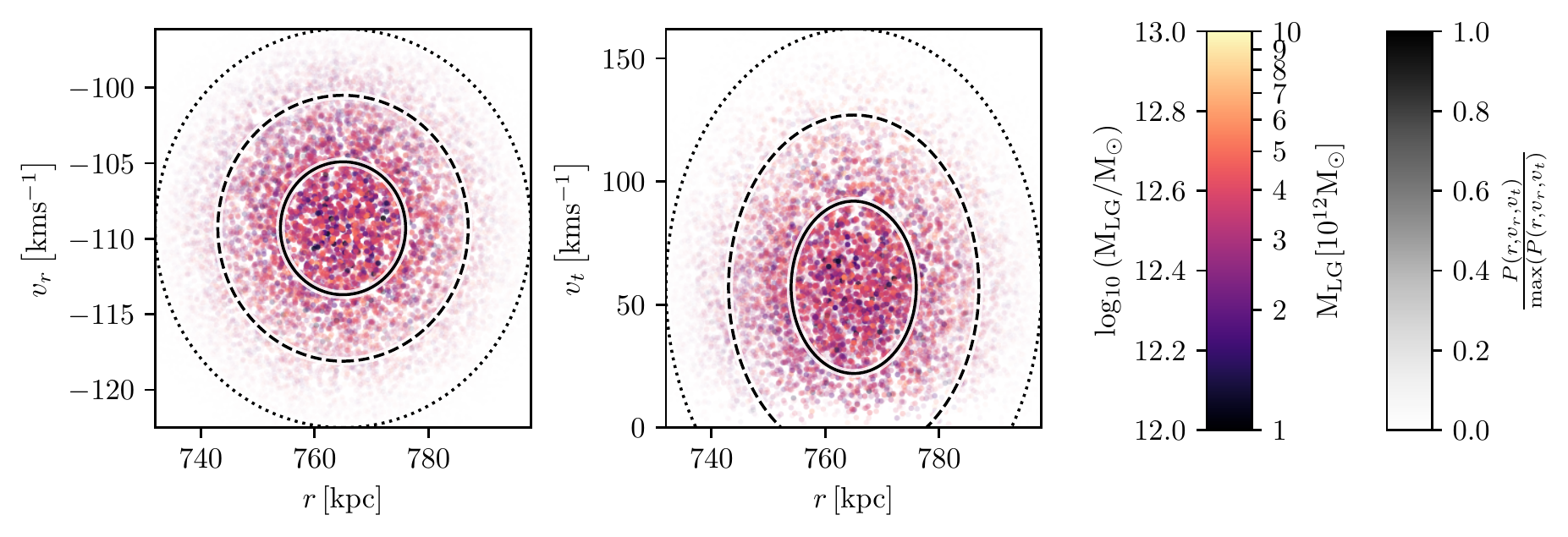}
    \vspace{-.3cm}
    \caption{LG analogues, coloured by mass, as a function of $r$ and $v_r$ (left column), and as a function of $r$ and $v_t$ (right column). The opacity of each data point (alpha value) is computed proportional to the probability that it represents the observed values of $r$, $v_r$, and $v_t$, computed using a 3D Gaussian kernel function (Equation~\ref{eqn:gaussian-distance}) in the ($r$, $v_r$, $v_t$)-space based on the observed values and their uncertainties. Although we include the full sample of tidally dominant systems with $\mathcal{F} < 0.5$ in our probability weighted estimate, for clarity, the axis ranges only extend out to $\pm 3 \sigma$ for each variable. For illustration, ellipses corresponding to $1 \sigma$, $2 \sigma$, and $3 \sigma$ in the two 2D-spaces ($r$, $v_r$) and ($r$, $v_t$) are overplotted. 
    }
    \label{fig:correlations-gaussian}
\end{figure*}

\begin{figure*}
    \includegraphics[width=7.1in] {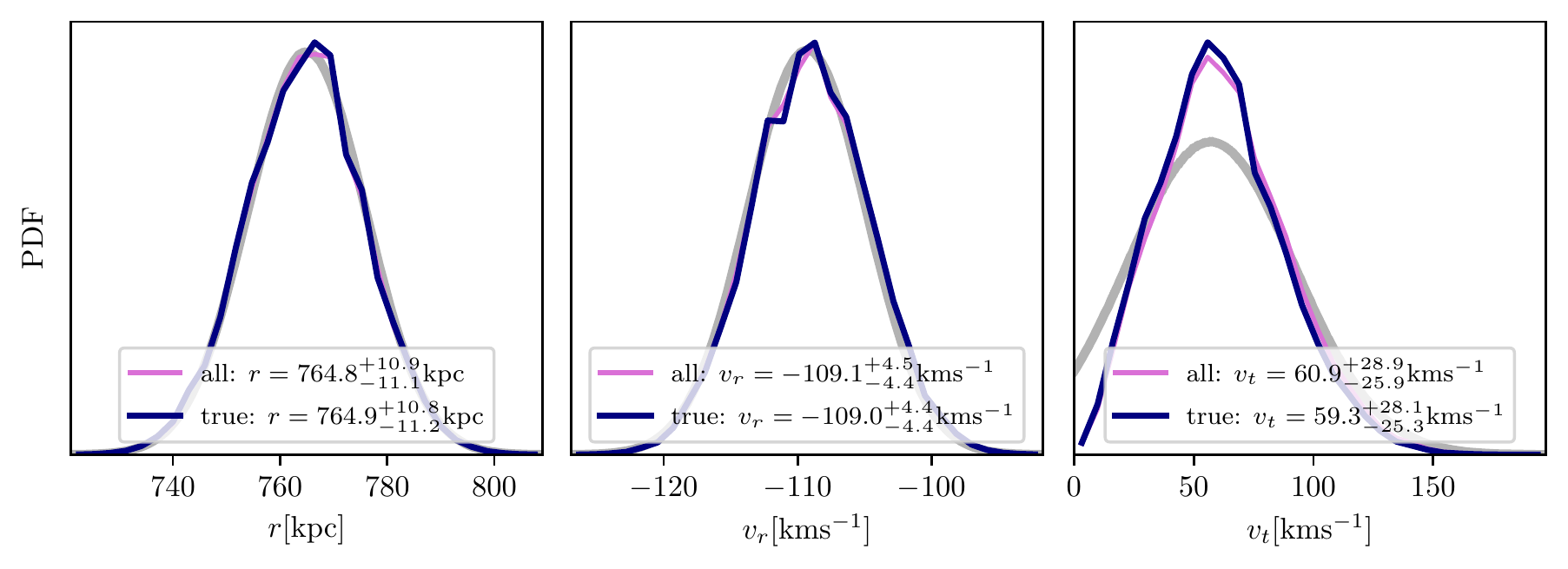}
    \vspace{-.5cm}
    \caption{Distributions of the kinematic variables $r$, $v_r$ and $v_t$ measured in the simulation with each LG analogue weighed according to its probability to represent the observed LG. On all panels, purple lines show the weighted estimate for the full set of LG analogues, blue lines show the weighted estimate for the `true' LG analogues that (tidally dominant and low force ratio). Grey lines indicate the assumed observational uncertainty of each variable. The distributions of $r$ and $v_r$, weighted by the overall probability, closely match the distributions assumed for the observations for each variable individually. The distributions of $v_t$, weighted by the overall probability, are less likely to contain values close to zero than the pure observational uncertainty would predict.
    }
    \label{fig:histograms-gaussian-features}
\end{figure*}

\section{Direct probability weighted estimates} \label{sec:results:weighted}
The uniquely large sample of LG analogues in {\sc Uchuu}, including many whose kinematics closely match the observations, also allows us to directly estimate the most likely mass of the Local Group and its members along with the associated uncertainties. For this purpose, we consider each LG analogue as a measurement of the observed LG, weighted according to a Gaussian kernel which describes the likelihood that it represents the observations.

For each LG analogue, $i$, we compute the distance to the observed kinematics as

\begin{eqnarray} \label{eqn:sigma}
\sigma_i = \sqrt{\sigma_{r,i}^2 + \sigma_{v_r,i}^2 + \sigma_{v_t,i}^2}
\end{eqnarray}
where $\sigma_{x,i} = (x_i - x_\mathrm{obs}) / \sigma_{x,\mathrm{obs}}$, with $x_\mathrm{obs}$ the mean observed value, and $\sigma_{x,\mathrm{obs}}$ the associated error, respectively, for each $x$ in $r, v_r, v_t$.

We then compute the relative probability that a LG analogue, $i$, represents the observations as

\begin{eqnarray}  \label{eqn:gaussian-distance}
    P_i = \frac{1}{(2 \pi)^{n/2}} e^{-\sigma_i^2/2}
\end{eqnarray}
where $n$ is the number of independent observables and $\sigma_i$ is as defined in Equation~\ref{eqn:sigma}. We use these probabilities as weights in computing the probability distributions of the posteriors given by the observables.

When we include additional constraints, such as the observed mass of the Milky Way or M31, we include these as additional observables, $x$, and compute the associated distances, $\sigma_i$, and probabilities, $P_i$, accordingly.

In Figure~\ref{fig:correlations-gaussian}, we show the distributions of the kinematic properties ($v_r$ and $r$ on the left, $v_t$ and $r$ on the right) of LG analogues, with the colour hue according to the LG mass (as in Figures~\ref{fig:correlations-tidal} and ~\ref{fig:correlations-force}, and the alpha-value, or opacity, according to the Gaussian kernel estimate of the relative probability that they represent the observed Local Group. In total, there are 2921 LG analogues with $\sigma(i) < 1$, 7129 with $\sigma(i) < 2$ and 11537 with $\sigma(i) < 3$, allowing us to make robust estimates of the distributions, not limited by sample size.

In Figure~\ref{fig:histograms-gaussian-features}, we show the histograms of the distributions of kinematic variables of LG analogues, weighed using the Gaussian distance in terms of $r$, $v_r$ and $v_t$. It can be seen that, for $r$ and $v_r$, both the median values and the standard deviations of the distributions are very close to the assumed observational values. This reflects the fact that the relative observational errors are so small that the distributions of $v_r$ and $r$ among the simulated LG analogues over these intervals are close to uniform. As discussed above, the observational uncertainty of $v_t$ is considerably larger, and given that $v_t$ is bounded from below by zero, the resulting distribution is asymmetric and skewed slightly towards higher values.

\begin{figure*}
    \includegraphics[width=7.1in]{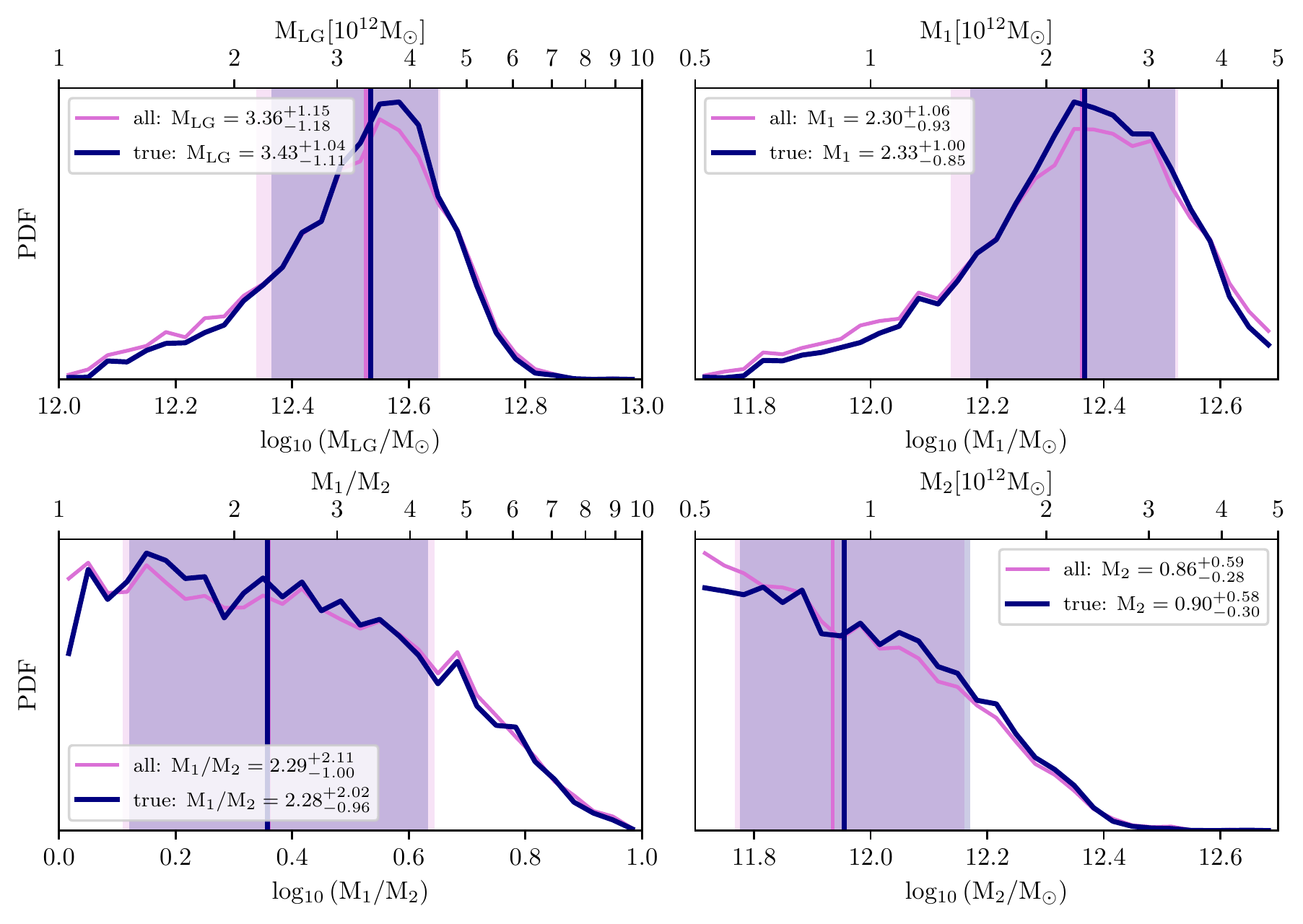}
    \vspace{-.5cm}
    \caption{Clockwise from the top left: posterior distributions of the total mass ($\mathrm{M_{LG}}$, top left), mass of the more massive halo ($\mathrm{M_{1}}$, top right), mass of the less massive halo ($\mathrm{M_{2}}$, bottom right) and mass ratio ($\mathrm{M_{M31} / M_{MW}}$, bottom left), when each LG analogue either from the full sample (purple lines) or the sample fulfilling the tidal dominance and force ratio criteria (dark blue lines) is weighted according to its likelihood to represent the observed LG, and assuming that M31 is the more massive of the pair. Shaded areas indicate the ranges between the 16$^{th}$ and 84$^{th}$ percentiles. Using this weighted sample, we obtain a median total mass of $3.43 \times 10^{12}\Ms$, and a median MW mass of $0.90 \times 10^{12} \Ms$, in line with the observations. We also obtain a median M31 mass of $2.33 \times 10^{12} \Ms$ and a median mass ratio of 2.28.
    }
    \label{fig:histograms-gaussian}
\end{figure*}

\subsection{Kinematically derived probability distributions}
In Figure~\ref{fig:histograms-gaussian}, we show the distributions of all four covariants, obtained using a probability weighted average: the total mass (top left), the mass ratio (top right), the lower of the two masses (bottom left) and the greater of the two masses (bottom right). On each panel, we show both the distribution for the full set of LG analogues, and for the subset of analogues that are both tidally dominant and with a low force ratio (`true pairs').

\begin{figure}
    \includegraphics[width=\columnwidth]{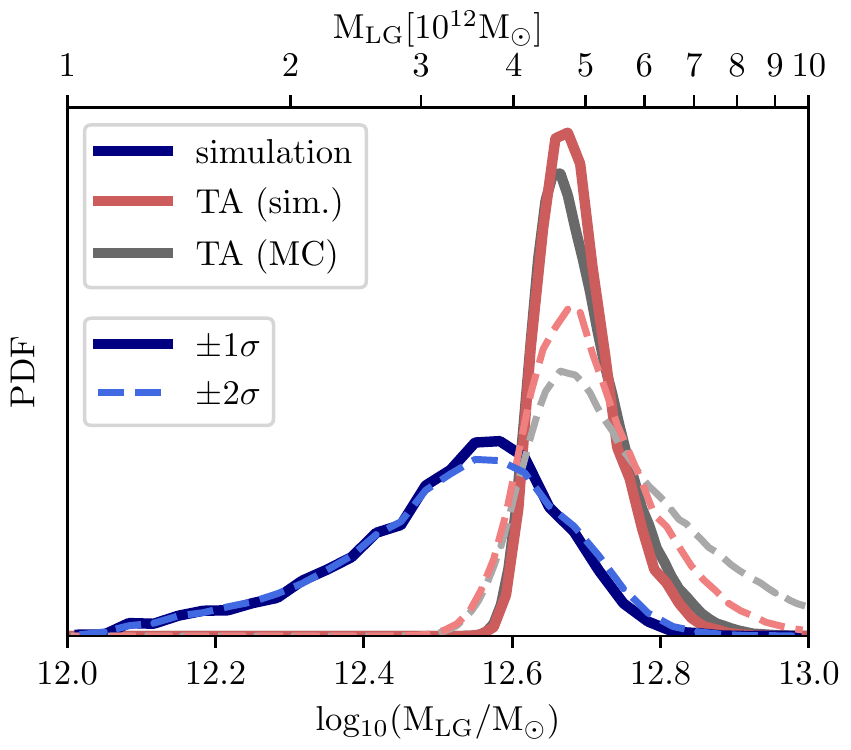}
    \vspace{-.3cm}
    \caption{Probability distribution of the LG mass. Blue, as measured in the simulation, weighted by the probability of representing the observations according to the assumed uncertainty. Red: as inferred from the timing argument for the same sample in the simulation with the same weights. Grey: as inferred from the timing argument, drawing Monte-Carlo samples with the same assumed means and uncertainties. Solid lines show the quoted $\pm 1 \sigma$ observational errors, dashed lines show the effect of increasing the assumed observational errors to $2 \sigma$. The precision of the timing argument estimate is limited only by the observational errors. By contrast, in the simulations, the uncertainty is dominated by the underlying scatter in the relation between mass and kinematics, imposing a fundamental limit on the precision.}
    \label{fig:TA-distributions}
\end{figure}

For these `true pairs', we find a total mass of $\mathrm{M_{LG}} = 3.45_{-1.12}^{+1.09} \times 10^{12} \Ms$. Both the centre of the distribution and the uncertainty, are very similar to the result obtained in Section~\ref{sec:results:regression} using Gaussian process regression. We find a mass of $\mathrm{M_2} = 0.9_{-0.30}^{+0.58}\times 10^{12} \Ms$ for the lower mass halo, and $\mathrm{M_1} = 2.33_{-0.85}^{+1.00}\times 10^{12} \Ms$ for the higher mass halo. 

The median total mass is considerably lower than most of the estimates using the timing argument. Notably, both individual masses and their uncertainties are consistent with the independent mass estimates for the MW and M31 listed discussed in Section~\ref{sec:previous}, at least if the MW is the less massive of the pair. We explore the mass-order in more detail in the next section.

It is also worth noting that the most likely (modal) mass of the lower mass halo, $M_2$, is near the lower limit of the mass range, $0.5 \times 10^{12} \Ms$. This is a consequence of the steep mass function in $\Lambda$CDM, and the fact that the kinematics constrain only the total mass, on which the lower mass halo has a comparatively small effect. As mentioned in Section~\ref{sec:selection}, the lower limit on $M_1$ that we selected must be considered as an additional prior.

\subsection{Comparison to the Timing Argument}
In Figure~\ref{fig:TA-distributions}, we show the probability distribution function for the probability-weighted estimate of $\mathrm{M_{LG}}$, the mass of the LG. The dark blue solid line shows the result we obtain using the $\pm 1 \sigma$ uncertainties described in Section~\ref{sec:selection}, $r=765 \pm 11$~kpc, $v_r=109.3 \pm 4.4$~kms$^{-1}$ and $v_t = 57 \pm 35$kms${-1}$. The light blue dashed line shows the result if we instead increase all observational errors by a factor of two:  $r=765 \pm 22$~kpc, $v_r=109.3 \pm 8.8$~kms$^{-1}$ and $v_t = 57 \pm 70$kms${-1}$. Confirming the results obtained through Gaussian Process regression discussed in Section~\ref{sec:results:regression}, the distribution function for the LG mass is almost unchanged. The true uncertainty in the mass of the LG is not limited by the observational errors in $r$, $v_r$ or $v_t$, but by the scatter in the kinematics-mass relation, i.e. the scatter in mass of objects with identical kinematics.

The other four lines in Figure~\ref{fig:TA-distributions} show the results of applying the timing argument, either to the kinematics of the probability weighted LG analogues in the simulation (red lines), or to Monte Carlo samples of the observations (grey lines). We compute timing argument masses by numerically solving the following four equations \citep{vanderMarel-2008}, where $a$ is the semi-major axis of the MW-M31 orbit, $e$ is the eccentricity, $\eta$ is the eccentric anomaly (which parameterises the orbital phase). The parameters $r$, $v_r$ and $v_t$ are the separation, radial, and transverse velocities, respectively, and $t_0$ is the age of the Universe.

\begin{eqnarray}
    r &=& a \left(1 - e \: \mathrm{cos} \: \eta\right), \\
    t_0 &=& \left(\frac{a^3}{GM}\right)^{1/2} \left(\eta - e \: \mathrm{sin} \: \eta\right), \\
    v_r &=& \left(\frac{GM}{a}\right)^{1/2} \frac{e \: \mathrm{sin} \: \eta} {1 - e \: \mathrm{cos} \: \eta}, \\
    v_t &=& \left(\frac{GM}{a}\right)^{1/2} \frac{(1 - e^2)^{1/2}}{1 - e \: \mathrm{cos} \: \eta}.
\end{eqnarray}

\begin{figure*}
    \includegraphics[width=7.1in]{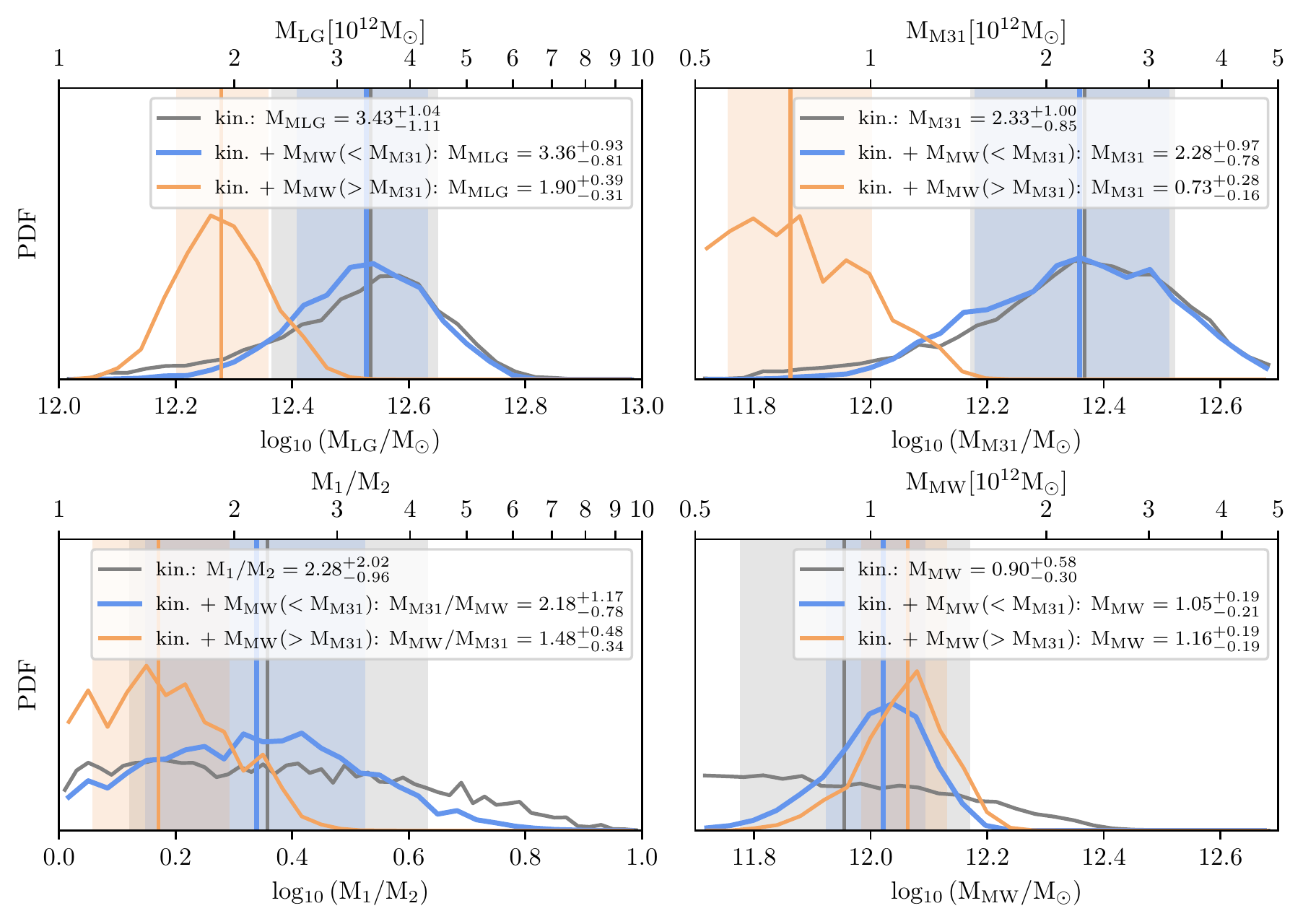}
    \vspace{-.5cm}
    \caption{Distributions of the total mass, mass of M31, mass of the MW, and the mass ratio, measured in the simulation with each LG analogue weighed according to its likelihood to represent the observed LG, either assuming only the observed kinematics (grey, as the dark blue lines in Figure~\ref{fig:histograms-gaussian}), or also assuming $1.1_{-0.2}^{+0.2} \times 10^{12}\Ms$ for the mass of the MW, as either the less massive of the pair ("regular mass order", in blue), or the more massive of the pair ("inverted mass order", in orange). Shaded areas show the regions between the 16$^{th}$ and 84$^{th}$ percentiles. Assuming that a MW of $\sim 1.1 \times 10^{12}\Ms$ is the less massive galaxy yields very similar results for the total mass, mass ratio, and mass of M31, with slightly reduced errors. By contrast, assuming a MW of $\sim 1.1 \times 10^{12}\Ms$ to be the more massive galaxy results in a much lower total mass, a much lower M31 mass, and a mass ratio much closer to unity.
    }
    \label{fig:histograms-gaussian-MW}
\end{figure*}

Note that the timing argument only applies to bound LG analogues and assumes that the LG is on its first approach. Both for the Monte Carlo samples and for the probability weighted kinematics, we also compute the distributions using either the quoted $\pm 1 \sigma$ uncertainty (solid lines), or increasing all assumed observational errors by a factor of two (dashed lines). By comparing the grey and red lines, we see that the probability weighted kinematics of the LG analogues in the simulations closely match the Monte Carlo samples, resulting in very similar distributions for the timing argument mass. By comparing the solid and dashed lines, we see that the precision of the timing argument estimate increases for smaller assumed errors. Indeed, if the observed errors tend to zero, the timing argument yields a unique solution. However, this does not reflect the true uncertainty in the relation between mass and kinematics measured in the simulations, as shown by the blue lines. We also note that the timing argument masses are offset from those measured directly in the simulations, suggesting that the timing argument is biassed.

\subsection{The Milky Way mass as an additional prior} \label{sec:results:MW}
Given the spate of recent measurements of the MW mass, we also investigate the dependence of Local Group properties when the MW mass is imposed as an additional constraint. The results are not purely Local Group kinematic estimates, and cannot be used directly to reveal (or alleviate) tensions between individual estimates and those obtained using pure kinematics. However, they may provide more accurate mass estimates, and a comparison to the purely kinematic results can also illuminate whether the different mass estimates are compatible.

As discussed in the introduction, the mass of the Milky Way is very well constrained to be close to $10^{12} \Ms$ using Gaia DR3 observations, among others. Assuming a value of $1.1_{-0.2}^{+0.2} \times 10^{12} \Ms$ as a fourth observable in Equations~\ref{eqn:sigma} and \ref{eqn:gaussian-distance}, we obtain new distributions for the masses of the individual objects, and for the mass of the Local Group, as shown in Figure~\ref{fig:histograms-gaussian-MW}. 

We consider both the case where the MW, at this mass, is the less massive or more massive halo of the pair. The first scenario, `regular mass order', is shown by the dark blue lines in Figure~\ref{fig:histograms-gaussian-MW}. Assuming  $M_{MW} = 1.1_{-0.2}^{+0.2} \times 10^{12} \Ms$ leads to a MW mass of $M_\mathrm{MW}=1.05_{-0.21}^{+0.19} \times 10^{12}\Ms$, an M31 mass of $M_\mathrm{M31}2.28_{+0.97}^{-0.78} \times 10^{12}\Ms$, a total mass of $3.36^{+0.93}_{-0.81} \times 10^{12}\Ms$, and a mass ratio of $2.18_{-0.78}^{+1.17}$. Reassuringly, all values are consistent with the values obtained purely from kinematics, showing that the kinematic analysis yields values compatible with the individual measurements of the MW mass. Using the measured MW mass as an additional observational constraint has increased the accuracy of the predictions for all three components. Under the assumption that the MW is the less massive of the pair, a MW mass of  $1.1_{-0.2}^{+0.2} \times 10^{12} $ is not in conflict with the LG kinematics. To the contrary, the MW mass is slightly above the median value inferred purely from kinematics.

When the MW is assumed to be the more massive of the two haloes (orange lines in Figure~\ref{fig:histograms-gaussian-MW}), the three other results change significantly. The total mass of the LG is reduced to $1.90^{+0.39}_{-0.31} \times 10^{12} \Ms$, while the mass of M31 is now $0.73_{-0.16}^{+0.28} \times 10^{12} \Ms$, and the mass ratio is $\mathrm{M_{MW} / M_{M31}} = 1.48_{-0.34}^{+0.48}$ ($\mathrm{M_{M31} / M_{MW}} = 0.68_{-0.17}^{+0.2}$). All these are outside the range predicted from the kinematics alone. The strong preference for an M31 mass below $10^{12} \Ms$ and a total mass below $2\times 10^{12} \Ms$ despite the kinematics in this `inverted mass ordering' stem from the fact that mass ratios very close to 1:1 are rare.

\begin{figure*}
    \includegraphics[width=7.1in]{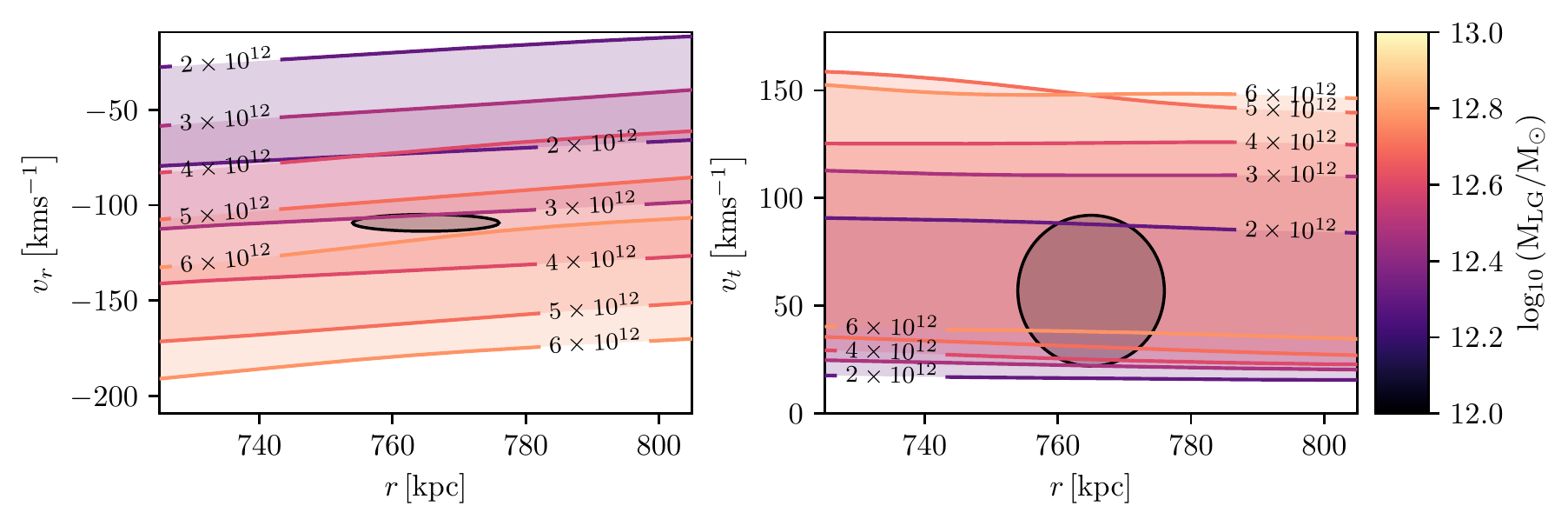}
    \vspace{-.5cm}
    \caption{Distributions of kinematic parameters for LG analogues with masses of $2 \pm 0.2\times10^{12}\Ms$, $3 \pm 0.2\times10^{12}\Ms$, $4 \pm 0.2\times10^{12}\Ms$, $5 \pm 0.2\times10^{12}\Ms$ and $6 \pm 0.2 \times10^{12}\Ms$. On both panels, contour lines correspond to $\pm 1 \sigma$ around the peak density, with the upper limits in $v_r$ and $v_t$ denoted on the left side, and the lower limits denoted on the right side of each panel. The ellipses on both panels show the observational uncertainty. In terms of $v_r$ and $r$, as shown on the left panel, the observed kinematics are typical for LG analogues with a mass of $\sim 4\times10^{12}\Ms$, and lie just outside the $1\sigma$ range for LG analogues with a mass of $3 \times10^{12}\Ms$ or $6 \times 10^{12} \Ms$. The scatter in $v_t$ is larger than that in $v_r$ in every mass range, and all masses are compatible with the measured $v_t$ value.}
    \label{fig:correlations-high-low}
\end{figure*}

The probability density functions shown in Figure~\ref{fig:histograms-gaussian-MW} show the estimated values of the masses and mass ratio given the assumed kinematics, and under the assumptions that a MW of  $1.1_{-0.2}^{+0.2} \times 10^{12}$ is either the less- or more massive halo of the pair. They do not, by themselves, predict which assumption for the mass ordering is more likely to be correct. To estimate the relative probability for a "regular" or "inverted" mass order given the observed kinematics and their associated uncertainties, we compare the sum of the weighted probability densities for both scenarios, which also accounts for the mass function and mass ratio distribution in the simulations. We find a ratio of 3.6, implying a $78\%$ chance that a MW of $1.1_{-0.2}^{+0.2} \times 10^{12}$ is less massive than M31.

\subsection{The inverse problem: possible kinematics for a given mass}\label{sec:results:inverse}
The kinematics strongly constrain the mass of the Local Group, but among the 1.4 million LG analogues, we find examples of Local Groups with any mass between $1-10 \times 10^{12} \Ms$ for nearly every combination of the kinematics. Part of the preference for a low-mass system is simply due to the fact that in\LCDM, low-mass haloes and by extension halo pairs, are far more prevalent than high mass haloes. It may therefore be illuminating to also consider the inverse question: given an assumed LG mass, what is the range of plausible kinematics?

In Figure~\ref{fig:correlations-high-low}, we show the distributions of the kinematics for LG analogues with masses of $\sim 2, 3, 4, 5$ and $6 \times10^{12} \Ms$. Each mass range is limited to $\pm 0.2 \times 10^{12}\Ms$, small enough to clearly separate different mass ranges, and large enough to include at least $10^4$ systems in each mass range. 

Also shown are ellipses centred on the observed values and extending to the $\pm 1 \sigma$ observational uncertainty. We find that the combination of separation, $r$, and transverse velocity $v_t$, does not sufficiently discriminate between different masses over this interval. Interestingly, this is not only due to the large observational error, but also due to the fact that the $1 \sigma$ scatter in $v_t$, ranges from $\sim 20-90$kms$^{-1}$ at $\mathrm{M} = 2 \times 10^{12} \Ms$ to $\sim 40-150$kms$^{-1}$ at $\mathrm{M} = 6 \times 10^{12} \Ms$. Each mass range includes a significant overlap with the observed value and uncertainty, and $v_t$ values as high as 125 kms$^{-1}$ are not uncommon for LG masses of up to $4 \times 10^{12} \Ms$. Higher mass pairs tend to have greater $v_t$ values, which in addition to the effects mentioned in Section~\ref{sec:results:kinematics} may partly explain the higher value reported by \cite{Forero-Romero-2022}, and values as high as $110$~kms$^{-1}$ can certainly not be ruled out. Only a value of $v_t$ as high as 164 kms$^{-1}$, as suggested by \cite{Salomon-2016}, is disfavoured (under 5\% probability in combination with the observations of $r$ and $v_r$) if the LG mass is not in excess of $4 \times 10^{12} \Ms$.

By contrast, the combination of separation, $r$, and radial velocity $v_r$, strongly depends on the mass. For LG analogues analogues with masses of $2 \times 10^{12}\Ms$ and a separation of $\sim 770$ kpc, the magnitude of the observed radial velocity is $2 \sigma$ above the median, while for LG analogues of $6 \times 10^{12}\Ms$, it is $1.2 \sigma$ below the median. The observed kinematics are most likely to occur for LG analogues of just under $4 \times 10^{12}\Ms$. However, the fact that low mass LGs are more common (and, to a smaller extent, the relatively low transverse velocity) pushes the most likely LG mass down to $\sim 3.3 \times 10^{12} \Ms$.

\section{Conclusion}\label{sec:conclusion}

We have inferred the halo masses of the Milky Way and M31, their combined mass, and mass ratio, by analysing kinematically selected LG analogues from the {\sc Uchuu} cosmological simulation. To establish the relation between mass and kinematics, we distinguish between `true pairs' in which the interactions between the MW and M31 dominate the gravitational force, and those whose kinematics are strongly affected by external influences. We find that this distinction significantly changes the distributions of kinematic variables.

Assuming \LCDM, limiting the range of halo masses to $0.5 - 5 \times 10^{12} \Ms$, and considering only true pairs where the gravitational field and tidal field are primarily due to the MW and M31 themselves, our analysis using Gaussian process regression results in a total mass of $\mathrm{M_{LG}} = 3.2_{-1.2}^{+0.9} \times 10^{12} \Ms$. We obtain a nearly identical result using a probability weighted estimate.

Our result for the mass of the LG  is significantly lower than most previous results which apply the timing argument to the kinematics, but in agreement with \cite{Benisty-2022}, who obtain a total mass of $3.4 \times 10^{12} \Ms$ when correcting the timing argument for cosmic bias. As discussed in Section~\ref{sec:results:inverse}, accounting for cosmic bias (the fact that there are more low mass than high mass haloes), which is folded into our analysis, is certainly important for determining the most likely mass. However, it is not clear that our agreement with \cite{Benisty-2022} is more than incidental, and we leave a more detailed analysis of the timing argument to future work.

We find that the kinematics most strongly constrain the total mass of the pair, leaving the mass of the two haloes and their mass ratio less well constrained. But importantly, the kinematics are not in conflict with the individual measurements. In particular, contrary to previous claims, they do not point towards a mass of the MW above $10^{12} \Ms$. In fact, a purely kinematic analysis gives a mass for the less massive halo of $0.9_{-0.31}^{+0.59} \times 10^{12} \Ms$. When we assume a MW mass of $1.1\pm0.2 \times 10^{12}\Ms$ as an additional prior, we also obtain essentially the same posteriors for the total mass, M31 mass, and mass ratio, assuming that the MW is the less massive of the pair. Lending strong support for the `regular' mass order, if we assume that the MW is more massive, all the distributions shift significantly, showing strong tensions with the observations.

We also find that LG analogues with close to equal masses are quite unlikely, and find a mass ratio of $2.28^{+2.02}_{-0.96}$ using only kinematics. It is worth noting that value and its distribution are quite sensitive to the lower bound of $5\times 10^{11} \Ms$ for each halo: allowing lower mass LG members is bound to shift the mass ratio distribution towards even higher values, while assuming a mass of $\sim 1.1 \pm 0.2 \times 10^{12} \Ms$ for the less massive halo gives a mass ratio of $2.18^{+1.17}_{-0.78} \Ms$. For the mass of the more massive halo, we find $2.33_{-0.85}^{+1.00} \times 10^{12} \Ms$ based only on pair kinematics, and we find $2.28_{-0.78}^{+0.97} \times 10^{12} \Ms$ for the mass of M31, assuming it is the more massive halo in a LG with a MW of $1.1 \pm 0.2 \times 10^{12} \Ms$.

It is worth noting that the pair kinematics only give a probability distribution for the mass of the LG. Both the Gaussian Process regression and our probability weighted estimate show significant tails towards higher and lower masses. To further constrain the mass (albeit forgoing the potential to reveal possible tensions), a combination with independent tracers such as the Hubble Flow or internal kinematics appears promising.

Importantly, the precision of our mass estimate is neither limited by statistics, nor by observational uncertainty. Instead, we find a fundamental and irreducible scatter in the relation between mass and kinematics of $\sim \pm 10^{12} \Ms$. This leads us to conclude that previous works which reported a smaller uncertainty purely based on the parameters of the Milky Way-Andromeda orbit, either through the timing argument or machine learning, did not fully account for this scatter. On the other hand, in order to obtain a more accurate mass estimate, future work will need to include additional information in the analysis.

\section*{Data Availability Statement}
This work is based entirely on public data. The {\sc Uchuu} simulation data is available for download at \url{http://skiesanduniverses.org/Simulations/Uchuu/} (See Acknowledgements).

Our analysis uses the open source software listed in the acknowledgements. Our documented analysis script, which can be used to reproduce all figures and in this paper from the above public data, is available in the form of documented python scripts and a jupyter notebook at \url{https://github.com/TillSawala/LG_Mass}.

\section*{Acknowledgements}
We thank Marius Cautun, Shihong Liao, Jorge Peñarubbia and Kai Puolamäki for insightful discussions and advice. We thank the creators of the {\sc Uchuu} simulation, \citeauthor{Ishiyama-2021}, for making their data publicly available.

TS acknowledges support from Academy of Finland grants 311049 and 339127, and from the European Research Council (ERC) Advanced Investigator grant DMIDAS (GA 786910). TS and MT were also supported by Academy of Finland grant 335607. PHJ acknowledges the support by the European Research Council via the ERC Consolidator Grant KETJU (no. 818930) and the Academy of Finland grant 339127.

This work used facilities hosted by the CSC - IT Centre for Science, Finland, and the DiRAC@Durham facility managed by the Institute for Computational Cosmology on behalf of the STFC DiRAC HPC Facility (www.dirac.ac.uk). The equipment was funded by BEIS capital funding via STFC capital grants ST/K00042X/1, ST/P002293/1, ST/R002371/1 and ST/S002502/1, Durham University and STFC operations grant ST/R000832/1. DiRAC is part of the National e-Infrastructure.

We thank Instituto de Astrofisica de Andalucia (IAA-CSIC), Centro de Supercomputacion de Galicia (CESGA) and the Spanish academic and research network (RedIRIS) in Spain for hosting {\sc Uchuu} DR1 and DR2 in the Skies \& Universes site for cosmological simulations. The {\sc Uchuu} simulations were carried out on Aterui II supercomputer at Center for Computational Astrophysics, CfCA, of National Astronomical Observatory of Japan, and the K computer at the RIKEN Advanced Institute for Computational Science. The {\sc Uchuu} DR1 and DR2 effort has made use of the skun@IAA RedIRIS and skun6@IAA computer facilities managed by the IAA-CSIC in Spain (MICINN EU-Feder grant EQC2018-004366-P).

We gratefully acknowledge the use of open source software, including Matplotlib \citep{matplotlib-paper}, scikit-learn \citep{scikit-learn}, SciPy \citep{SciPy} and NumPy \citep{numpy-paper}.

%%%%%%%%%%%%%%%%%%%%%%%%%%%%%%%%%%%%%%%%%%%%%%%%%%


\bibliographystyle{mnras} \bibliography{paper} 

\begin{thebibliography}{}
\makeatletter
\relax
\def\mn@urlcharsother{\let\do\@makeother \do\$\do\&\do\#\do\^\do\_\do\%\do\~}
\def\mn@doi{\begingroup\mn@urlcharsother \@ifnextchar [ {\mn@doi@}
  {\mn@doi@[]}}
\def\mn@doi@[#1]#2{\def\@tempa{#1}\ifx\@tempa\@empty \href
  {http://dx.doi.org/#2} {doi:#2}\else \href {http://dx.doi.org/#2} {#1}\fi
  \endgroup}
\def\mn@eprint#1#2{\mn@eprint@#1:#2::\@nil}
\def\mn@eprint@arXiv#1{\href {http://arxiv.org/abs/#1} {{\tt arXiv:#1}}}
\def\mn@eprint@dblp#1{\href {http://dblp.uni-trier.de/rec/bibtex/#1.xml}
  {dblp:#1}}
\def\mn@eprint@#1:#2:#3:#4\@nil{\def\@tempa {#1}\def\@tempb {#2}\def\@tempc
  {#3}\ifx \@tempc \@empty \let \@tempc \@tempb \let \@tempb \@tempa \fi \ifx
  \@tempb \@empty \def\@tempb {arXiv}\fi \@ifundefined
  {mn@eprint@\@tempb}{\@tempb:\@tempc}{\expandafter \expandafter \csname
  mn@eprint@\@tempb\endcsname \expandafter{\@tempc}}}

\bibitem[\protect\citeauthoryear{{Ablimit}, {Zhao}, {Flynn}  \&
  {Bird}}{{Ablimit} et~al.}{2020}]{Ablimit-2020}
{Ablimit} I.,  {Zhao} G.,  {Flynn} C.,   {Bird} S.~A.,  2020, \mn@doi [\apjl]
  {10.3847/2041-8213/ab8d45}, \href
  {https://ui.adsabs.harvard.edu/abs/2020ApJ...895L..12A} {895, L12}

\bibitem[\protect\citeauthoryear{{Benisty}, {Vasiliev}, {Evans}, {Davis},
  {Hartl}  \& {Strigari}}{{Benisty} et~al.}{2022}]{Benisty-2022}
{Benisty} D.,  {Vasiliev} E.,  {Evans} N.~W.,  {Davis} A.-C.,  {Hartl} O.~V.,
  {Strigari} L.~E.,  2022, \mn@doi [\apjl] {10.3847/2041-8213/ac5c42}, \href
  {https://ui.adsabs.harvard.edu/abs/2022ApJ...928L...5B} {928, L5}

\bibitem[\protect\citeauthoryear{{Boylan-Kolchin}, {Bullock}  \&
  {Kaplinghat}}{{Boylan-Kolchin} et~al.}{2012}]{Boylan-Kolchin-2012}
{Boylan-Kolchin} M.,  {Bullock} J.~S.,   {Kaplinghat} M.,  2012, \mn@doi
  [\mnras] {10.1111/j.1365-2966.2012.20695.x}, \href
  {http://adsabs.harvard.edu/abs/2012MNRAS.422.1203B} {422, 1203}

\bibitem[\protect\citeauthoryear{{Callingham} et~al.,}{{Callingham}
  et~al.}{2019}]{Callingham-2019}
{Callingham} T.~M.,  et~al., 2019, \mn@doi [\mnras] {10.1093/mnras/stz365},
  \href {https://ui.adsabs.harvard.edu/abs/2019MNRAS.484.5453C} {484, 5453}

\bibitem[\protect\citeauthoryear{Carlesi et~al.,}{Carlesi
  et~al.}{2016}]{Carlesi-2016}
Carlesi E.,  et~al., 2016, \mn@doi [Monthly Notices of the Royal Astronomical
  Society] {10.1093/mnras/stw357}, 458, 900–911

\bibitem[\protect\citeauthoryear{{Cautun} et~al.,}{{Cautun}
  et~al.}{2020}]{Cautun-2020}
{Cautun} M.,  et~al., 2020, \mn@doi [\mnras] {10.1093/mnras/staa1017}, \href
  {https://ui.adsabs.harvard.edu/abs/2020MNRAS.494.4291C} {494, 4291}

\bibitem[\protect\citeauthoryear{{Chamberlain}, {Price-Whelan}, {Besla},
  {Cunningham}, {Garavito-Camargo}, {Pe{\~n}arrubia}  \&
  {Petersen}}{{Chamberlain} et~al.}{2022}]{Chamberlain-2022}
{Chamberlain} K.,  {Price-Whelan} A.~M.,  {Besla} G.,  {Cunningham} E.~C.,
  {Garavito-Camargo} N.,  {Pe{\~n}arrubia} J.,   {Petersen} M.~S.,  2022, arXiv
  e-prints, \href {https://ui.adsabs.harvard.edu/abs/2022arXiv220407173C} {p.
  arXiv:2204.07173}

\bibitem[\protect\citeauthoryear{{Deason} et~al.,}{{Deason}
  et~al.}{2021}]{Deason-2021}
{Deason} A.~J.,  et~al., 2021, \mn@doi [\mnras] {10.1093/mnras/staa3984}, \href
  {https://ui.adsabs.harvard.edu/abs/2021MNRAS.501.5964D} {501, 5964}

\bibitem[\protect\citeauthoryear{{Fardal} et~al.,}{{Fardal}
  et~al.}{2013}]{Fardal-2013}
{Fardal} M.~A.,  et~al., 2013, \mn@doi [\mnras] {10.1093/mnras/stt1121}, \href
  {https://ui.adsabs.harvard.edu/abs/2013MNRAS.434.2779F} {434, 2779}

\bibitem[\protect\citeauthoryear{{Fattahi} et~al.,}{{Fattahi}
  et~al.}{2015}]{Fattahi-2015}
{Fattahi} A.,  et~al., 2015, ArXiv 1507.03643, \href
  {http://adsabs.harvard.edu/abs/2015arXiv150703643F} {}

\bibitem[\protect\citeauthoryear{{Font}, {McCarthy}, {Belokurov}, {Brown}  \&
  {Stafford}}{{Font} et~al.}{2022}]{Font-2022}
{Font} A.~S.,  {McCarthy} I.~G.,  {Belokurov} V.,  {Brown} S.~T.,   {Stafford}
  S.~G.,  2022, \mn@doi [\mnras] {10.1093/mnras/stac183}, \href
  {https://ui.adsabs.harvard.edu/abs/2022MNRAS.511.1544F} {511, 1544}

\bibitem[\protect\citeauthoryear{{Forero-Romero} \&
  {Sierra-Porta}}{{Forero-Romero} \& {Sierra-Porta}}{2022}]{Forero-Romero-2022}
{Forero-Romero} J.~E.,  {Sierra-Porta} D.,  2022, arXiv e-prints, \href
  {https://ui.adsabs.harvard.edu/abs/2022arXiv220909369F} {p. arXiv:2209.09369}

\bibitem[\protect\citeauthoryear{{Fritz}, {Di Cintio}, {Battaglia}, {Brook}  \&
  {Taibi}}{{Fritz} et~al.}{2020}]{Fritz-2020}
{Fritz} T.~K.,  {Di Cintio} A.,  {Battaglia} G.,  {Brook} C.,   {Taibi} S.,
  2020, \mn@doi [\mnras] {10.1093/mnras/staa1040}, \href
  {https://ui.adsabs.harvard.edu/abs/2020MNRAS.494.5178F} {494, 5178}

\bibitem[\protect\citeauthoryear{{Guo}, {White}, {Li}  \&
  {Boylan-Kolchin}}{{Guo} et~al.}{2010}]{Guo-2010}
{Guo} Q.,  {White} S.,  {Li} C.,   {Boylan-Kolchin} M.,  2010, \mn@doi [\mnras]
  {10.1111/j.1365-2966.2010.16341.x}, \href
  {http://adsabs.harvard.edu/abs/2010MNRAS.404.1111G} {404, 1111}

\bibitem[\protect\citeauthoryear{{Guo} et~al.,}{{Guo} et~al.}{2011}]{Guo-2011}
{Guo} Q.,  et~al., 2011, \mn@doi [\mnras] {10.1111/j.1365-2966.2010.18114.x},
  \href {http://adsabs.harvard.edu/abs/2011MNRAS.413..101G} {413, 101}

\bibitem[\protect\citeauthoryear{Harris et~al.,}{Harris
  et~al.}{2020}]{numpy-paper}
Harris C.~R.,  et~al., 2020, \mn@doi [Nature] {10.1038/s41586-020-2649-2}, 585,
  357

\bibitem[\protect\citeauthoryear{{Hartl} \& {Strigari}}{{Hartl} \&
  {Strigari}}{2022}]{Hartl-2022}
{Hartl} O.~V.,  {Strigari} L.~E.,  2022, \mn@doi [\mnras]
  {10.1093/mnras/stac413}, \href
  {https://ui.adsabs.harvard.edu/abs/2022MNRAS.511.6193H} {511, 6193}

\bibitem[\protect\citeauthoryear{Hunter}{Hunter}{2007}]{matplotlib-paper}
Hunter J.~D.,  2007, \mn@doi [Computing in Science \& Engineering]
  {10.1109/MCSE.2007.55}, 9, 90

\bibitem[\protect\citeauthoryear{{Ishiyama} et~al.,}{{Ishiyama}
  et~al.}{2021}]{Ishiyama-2021}
{Ishiyama} T.,  et~al., 2021, \mn@doi [\mnras] {10.1093/mnras/stab1755}, \href
  {https://ui.adsabs.harvard.edu/abs/2021MNRAS.506.4210I} {506, 4210}

\bibitem[\protect\citeauthoryear{{Kafle}, {Sharma}, {Lewis}, {Robotham}  \&
  {Driver}}{{Kafle} et~al.}{2018}]{Kafle-2018}
{Kafle} P.~R.,  {Sharma} S.,  {Lewis} G.~F.,  {Robotham} A. S.~G.,   {Driver}
  S.~P.,  2018, \mn@doi [\mnras] {10.1093/mnras/sty082}, \href
  {https://ui.adsabs.harvard.edu/abs/2018MNRAS.475.4043K} {475, 4043}

\bibitem[\protect\citeauthoryear{{Karukes}, {Benito}, {Iocco}, {Trotta}  \&
  {Geringer-Sameth}}{{Karukes} et~al.}{2020}]{Karukes-2020}
{Karukes} E.~V.,  {Benito} M.,  {Iocco} F.,  {Trotta} R.,   {Geringer-Sameth}
  A.,  2020, \mn@doi [\jcap] {10.1088/1475-7516/2020/05/033}, \href
  {https://ui.adsabs.harvard.edu/abs/2020JCAP...05..033K} {2020, 033}

\bibitem[\protect\citeauthoryear{{Kennedy}, {Frenk}, {Cole}  \&
  {Benson}}{{Kennedy} et~al.}{2014}]{Kennedy-2014}
{Kennedy} R.,  {Frenk} C.,  {Cole} S.,   {Benson} A.,  2014, \mn@doi [\mnras]
  {10.1093/mnras/stu719}, \href
  {http://adsabs.harvard.edu/abs/2014MNRAS.442.2487K} {442, 2487}

\bibitem[\protect\citeauthoryear{{Li} \& {White}}{{Li} \&
  {White}}{2008}]{Li-2008}
{Li} Y.-S.,  {White} S.~D.~M.,  2008, \mn@doi [\mnras]
  {10.1111/j.1365-2966.2007.12748.x}, \href
  {http://adsabs.harvard.edu/abs/2008MNRAS.384.1459L} {384, 1459}

\bibitem[\protect\citeauthoryear{{Li}, {Qian}, {Han}, {Li}, {Wang}  \&
  {Jing}}{{Li} et~al.}{2020}]{Li-2020}
{Li} Z.-Z.,  {Qian} Y.-Z.,  {Han} J.,  {Li} T.~S.,  {Wang} W.,   {Jing} Y.~P.,
  2020, \mn@doi [\apj] {10.3847/1538-4357/ab84f0}, \href
  {https://ui.adsabs.harvard.edu/abs/2020ApJ...894...10L} {894, 10}

\bibitem[\protect\citeauthoryear{{Libeskind}, {Yepes}, {Knebe},
  {Gottl{\"o}ber}, {Hoffman}  \& {Knollmann}}{{Libeskind}
  et~al.}{2010}]{Libeskind-2010}
{Libeskind} N.~I.,  {Yepes} G.,  {Knebe} A.,  {Gottl{\"o}ber} S.,  {Hoffman}
  Y.,   {Knollmann} S.~R.,  2010, \mn@doi [\mnras]
  {10.1111/j.1365-2966.2009.15766.x}, \href
  {https://ui.adsabs.harvard.edu/abs/2010MNRAS.401.1889L} {401, 1889}

\bibitem[\protect\citeauthoryear{{Lovell}, {Frenk}, {Eke}, {Jenkins}, {Gao}  \&
  {Theuns}}{{Lovell} et~al.}{2014}]{Lovell-2014}
{Lovell} M.~R.,  {Frenk} C.~S.,  {Eke} V.~R.,  {Jenkins} A.,  {Gao} L.,
  {Theuns} T.,  2014, \mn@doi [\mnras] {10.1093/mnras/stt2431}, \href
  {https://ui.adsabs.harvard.edu/abs/2014MNRAS.439..300L} {439, 300}

\bibitem[\protect\citeauthoryear{{McLeod}, {Libeskind}, {Lahav}  \&
  {Hoffman}}{{McLeod} et~al.}{2017}]{McLeod-2017}
{McLeod} M.,  {Libeskind} N.,  {Lahav} O.,   {Hoffman} Y.,  2017, \mn@doi
  [\jcap] {10.1088/1475-7516/2017/12/034}, \href
  {https://ui.adsabs.harvard.edu/abs/2017JCAP...12..034M} {2017, 034}

\bibitem[\protect\citeauthoryear{{Partridge}, {Lahav}  \&
  {Hoffman}}{{Partridge} et~al.}{2013}]{Partridge-2013}
{Partridge} C.,  {Lahav} O.,   {Hoffman} Y.,  2013, \mn@doi [\mnras]
  {10.1093/mnrasl/slt109}, \href
  {https://ui.adsabs.harvard.edu/abs/2013MNRAS.436L..45P} {436, L45}

\bibitem[\protect\citeauthoryear{{Pawlowski}}{{Pawlowski}}{2018}]{Pawlowski-review}
{Pawlowski} M.~S.,  2018, \mn@doi [Modern Physics Letters A]
  {10.1142/S0217732318300045}, \href
  {https://ui.adsabs.harvard.edu/abs/2018MPLA...3330004P} {33, 1830004}

\bibitem[\protect\citeauthoryear{{Pe{\~n}arrubia}, {Ma}, {Walker}  \&
  {McConnachie}}{{Pe{\~n}arrubia} et~al.}{2014}]{Penarrubia-2014}
{Pe{\~n}arrubia} J.,  {Ma} Y.-Z.,  {Walker} M.~G.,   {McConnachie} A.,  2014,
  \mn@doi [\mnras] {10.1093/mnras/stu879}, \href
  {http://adsabs.harvard.edu/abs/2014MNRAS.443.2204P} {443, 2204}

\bibitem[\protect\citeauthoryear{Pedregosa et~al.,}{Pedregosa
  et~al.}{2011}]{scikit-learn}
Pedregosa F.,  et~al., 2011, Journal of Machine Learning Research, 12, 2825

\bibitem[\protect\citeauthoryear{{Petersen} \& {Pe{\~n}arrubia}}{{Petersen} \&
  {Pe{\~n}arrubia}}{2020}]{Petersen-2020}
{Petersen} M.~S.,  {Pe{\~n}arrubia} J.,  2020, \mn@doi [\mnras]
  {10.1093/mnrasl/slaa029}, \href
  {https://ui.adsabs.harvard.edu/abs/2020MNRAS.494L..11P} {494, L11}

\bibitem[\protect\citeauthoryear{Peñarrubia \& Fattahi}{Peñarrubia \&
  Fattahi}{2017}]{Penarrubia-2017}
Peñarrubia J.,  Fattahi A.,  2017, \mn@doi [Monthly Notices of the Royal
  Astronomical Society] {10.1093/mnras/stx323}, 468, 1300–1316

\bibitem[\protect\citeauthoryear{{Prudil} et~al.,}{{Prudil}
  et~al.}{2022}]{Prudil-2022}
{Prudil} Z.,  et~al., 2022, arXiv e-prints, \href
  {https://ui.adsabs.harvard.edu/abs/2022arXiv220600417P} {p. arXiv:2206.00417}

\bibitem[\protect\citeauthoryear{{Rodriguez Wimberly} et~al.,}{{Rodriguez
  Wimberly} et~al.}{2022}]{Rodriguez-Wimberly-2022}
{Rodriguez Wimberly} M.~K.,  et~al., 2022, \mn@doi [\mnras]
  {10.1093/mnras/stac1265}, \href
  {https://ui.adsabs.harvard.edu/abs/2022MNRAS.513.4968R} {513, 4968}

\bibitem[\protect\citeauthoryear{{Salomon}, {Ibata}, {Famaey}, {Martin}  \&
  {Lewis}}{{Salomon} et~al.}{2016}]{Salomon-2016}
{Salomon} J.~B.,  {Ibata} R.~A.,  {Famaey} B.,  {Martin} N.~F.,   {Lewis}
  G.~F.,  2016, \mn@doi [\mnras] {10.1093/mnras/stv2865}, \href
  {https://ui.adsabs.harvard.edu/abs/2016MNRAS.456.4432S} {456, 4432}

\bibitem[\protect\citeauthoryear{{Salomon}, {Ibata}, {Reyl{\'e}}, {Famaey},
  {Libeskind}, {McConnachie}  \& {Hoffman}}{{Salomon}
  et~al.}{2020}]{Salomon-2020}
{Salomon} J.~B.,  {Ibata} R.,  {Reyl{\'e}} C.,  {Famaey} B.,  {Libeskind}
  N.~I.,  {McConnachie} A.~W.,   {Hoffman} Y.,  2020, arXiv e-prints, \href
  {https://ui.adsabs.harvard.edu/abs/2020arXiv201209204S} {p. arXiv:2012.09204}

\bibitem[\protect\citeauthoryear{{Sawala} et~al.,}{{Sawala}
  et~al.}{2015}]{Sawala-2015}
{Sawala} T.,  et~al., 2015, \mn@doi [\mnras] {10.1093/mnras/stu2753}, \href
  {http://adsabs.harvard.edu/abs/2015MNRAS.448.2941S} {448, 2941}

\bibitem[\protect\citeauthoryear{{Sawala} et~al.,}{{Sawala}
  et~al.}{2022}]{Sawala-2022}
{Sawala} T.,  et~al., 2022, arXiv e-prints, \href
  {https://ui.adsabs.harvard.edu/abs/2022arXiv220502860S} {p. arXiv:2205.02860}

\bibitem[\protect\citeauthoryear{{Shen} et~al.,}{{Shen}
  et~al.}{2022}]{Shen-2022}
{Shen} J.,  et~al., 2022, \mn@doi [\apj] {10.3847/1538-4357/ac3a7a}, \href
  {https://ui.adsabs.harvard.edu/abs/2022ApJ...925....1S} {925, 1}

\bibitem[\protect\citeauthoryear{{Slizewski}, {Dufresne}, {Murdock}, {Eadie},
  {Sanderson}, {Wetzel}  \& {Juri{\'c}}}{{Slizewski}
  et~al.}{2022}]{Slizewski-2022}
{Slizewski} A.,  {Dufresne} X.,  {Murdock} K.,  {Eadie} G.,  {Sanderson} R.,
  {Wetzel} A.,   {Juri{\'c}} M.,  2022, \mn@doi [\apj]
  {10.3847/1538-4357/ac390b}, \href
  {https://ui.adsabs.harvard.edu/abs/2022ApJ...924..131S} {924, 131}

\bibitem[\protect\citeauthoryear{{Tamm}, {Tempel}, {Tenjes}, {Tihhonova}  \&
  {Tuvikene}}{{Tamm} et~al.}{2012}]{Tamm-2012}
{Tamm} A.,  {Tempel} E.,  {Tenjes} P.,  {Tihhonova} O.,   {Tuvikene} T.,  2012,
  \mn@doi [\aap] {10.1051/0004-6361/201220065}, \href
  {http://adsabs.harvard.edu/abs/2012A%26A...546A...4T} {546, A4}

\bibitem[\protect\citeauthoryear{{Tollerud} et~al.,}{{Tollerud}
  et~al.}{2012}]{Tollerud-2012}
{Tollerud} E.~J.,  et~al., 2012, \mn@doi [\apj] {10.1088/0004-637X/752/1/45},
  \href {https://ui.adsabs.harvard.edu/abs/2012ApJ...752...45T} {752, 45}

\bibitem[\protect\citeauthoryear{{Veljanoski} et~al.,}{{Veljanoski}
  et~al.}{2013}]{Veljanoski-2013}
{Veljanoski} J.,  et~al., 2013, \mn@doi [\apjl] {10.1088/2041-8205/768/2/L33},
  \href {https://ui.adsabs.harvard.edu/abs/2013ApJ...768L..33V} {768, L33}

\bibitem[\protect\citeauthoryear{{Veljanoski} et~al.,}{{Veljanoski}
  et~al.}{2014}]{Veljanoski-2014}
{Veljanoski} J.,  et~al., 2014, \mn@doi [\mnras] {10.1093/mnras/stu1055}, \href
  {https://ui.adsabs.harvard.edu/abs/2014MNRAS.442.2929V} {442, 2929}

\bibitem[\protect\citeauthoryear{Virtanen et~al.,}{Virtanen
  et~al.}{2020}]{SciPy}
Virtanen P.,  et~al., 2020, \mn@doi [Nature Methods]
  {10.1038/s41592-019-0686-2}, \href {https://rdcu.be/b08Wh} {17, 261}

\bibitem[\protect\citeauthoryear{{Wang}}{{Wang}}{2020}]{Wang-2020-ML}
{Wang} J.,  2020, arXiv e-prints, \href
  {https://ui.adsabs.harvard.edu/abs/2020arXiv200910862W} {p. arXiv:2009.10862}

\bibitem[\protect\citeauthoryear{{Wang}, {Han}, {Cautun}, {Li}  \&
  {Ishigaki}}{{Wang} et~al.}{2020}]{Wang-2020}
{Wang} W.,  {Han} J.,  {Cautun} M.,  {Li} Z.,   {Ishigaki} M.~N.,  2020,
  \mn@doi [Science China Physics, Mechanics, and Astronomy]
  {10.1007/s11433-019-1541-6}, \href
  {https://ui.adsabs.harvard.edu/abs/2020SCPMA..6309801W} {63, 109801}

\bibitem[\protect\citeauthoryear{{Watkins}, {Evans}  \& {An}}{{Watkins}
  et~al.}{2010}]{Watkins-2010}
{Watkins} L.~L.,  {Evans} N.~W.,   {An} J.~H.,  2010, \mn@doi [\mnras]
  {10.1111/j.1365-2966.2010.16708.x}, \href
  {https://ui.adsabs.harvard.edu/abs/2010MNRAS.406..264W} {406, 264}

\bibitem[\protect\citeauthoryear{{Watkins}, {van der Marel}, {Sohn}  \&
  {Evans}}{{Watkins} et~al.}{2019}]{Watkins-2019}
{Watkins} L.~L.,  {van der Marel} R.~P.,  {Sohn} S.~T.,   {Evans} N.~W.,  2019,
  \mn@doi [\apj] {10.3847/1538-4357/ab089f}, \href
  {https://ui.adsabs.harvard.edu/abs/2019ApJ...873..118W} {873, 118}

\bibitem[\protect\citeauthoryear{van~der Marel \& Guhathakurta}{van~der Marel
  \& Guhathakurta}{2008}]{vanderMarel-2008}
van~der Marel R.~P.,  Guhathakurta P.,  2008, \mn@doi [The Astrophysical
  Journal] {10.1086/533430}, 678, 187–199

\bibitem[\protect\citeauthoryear{{van der Marel}, {Fardal}, {Besla}, {Beaton},
  {Sohn}, {Anderson}, {Brown}  \& {Guhathakurta}}{{van der Marel}
  et~al.}{2012}]{vanderMarel-2012}
{van der Marel} R.~P.,  {Fardal} M.,  {Besla} G.,  {Beaton} R.~L.,  {Sohn}
  S.~T.,  {Anderson} J.,  {Brown} T.,   {Guhathakurta} P.,  2012, \mn@doi
  [\apj] {10.1088/0004-637X/753/1/8}, \href
  {http://adsabs.harvard.edu/abs/2012ApJ...753....8V} {753, 8}

\bibitem[\protect\citeauthoryear{{van der Marel}, {Fardal}, {Sohn}, {Patel},
  {Besla}, {del Pino}, {Sahlmann}  \& {Watkins}}{{van der Marel}
  et~al.}{2019}]{vanderMarel-2019}
{van der Marel} R.~P.,  {Fardal} M.~A.,  {Sohn} S.~T.,  {Patel} E.,  {Besla}
  G.,  {del Pino} A.,  {Sahlmann} J.,   {Watkins} L.~L.,  2019, \mn@doi [\apj]
  {10.3847/1538-4357/ab001b}, \href
  {https://ui.adsabs.harvard.edu/abs/2019ApJ...872...24V} {872, 24}

\makeatother
\end{thebibliography}

%%%%%%%%%%%%%%%%%%%%%%%%%%%%%%%%%%%%%%%%%%%%%%%%%%

%%%%%%%%%%%%%%%%% APPENDICES %%%%%%%%%%%%%%%%%%%%%
\bsp	% typesetting comment
\label{lastpage}
\end{document}